\documentclass[11pt,twocolumn]{article}
\usepackage{authblk}

\usepackage[utf8]{inputenc}
\usepackage{graphicx}
\usepackage[a4paper]{geometry}
\title{Maximizing the electromagnetic efficiency of spintronic terahertz emitters}

\author[1,2]{Pierre Koleják}
\author[2]{Geoffrey Lezier}
\author[1]{Daniel Vala}
\author[2]{Baptiste Mathmann}
\author[1]{Lukáš Halagačka}
\author[1]{Zuzana Gelnárová}
\author[2]{Yannick Dusch}
\author[1]{Jean-Fran\c{c}ois Lampin}
\author[2]{Nicolas Tiercelin}
\author[1]{Kamil Postava}
\author[2]{Mathias Vanwolleghem}
\affil[1]{IT4Innovations National Supercomputing Center \& Faculty of Materials Science and Technology, VSB -- Technical University of Ostrava, 17.\,listopadu 15, 708\,00 Ostrava, Czech Republic}
\affil[2]{ Université de Lille, CNRS, Centrale Lille, Université Polytechnique Hauts-de-France, UMR 8520 - IEMN, 59000 Lille, France}
\date{}

\usepackage{float}
\usepackage{hyperref}
\usepackage[dvipsnames]{xcolor}
\usepackage{amsmath, calc}
\usepackage[normalem]{ulem}
\usepackage{csquotes}
\usepackage{todonotes}

\usepackage{adjustbox}
\usepackage{comment}

\begin{document}
\maketitle
\sloppy


\section*{Abstract}

Spintronic Terahertz Emitters (STE) represent a significant advancement in source technology, exploiting the ultrafast demagnetization process of spin-electrons to unveil a 30 THz wide, gapless spectrum, accessible through femtosecond lasers across the full VIS-IR range. This innovation not only positions STEs as a pivotal advancement in terahertz source technology but also underscores their role as a cost-effective, high-performance solution, thereby redefining standards within the field. However, the inherent spintronic nature of these devices introduces a challenge: a lower optical-to-terahertz conversion efficiency, which positions them at a notable disadvantage relative to other sources.
In response, this work aims to substantially improve the electromagnetic efficiency of these emitters. This is accomplished by maximizing the energy conversion from the pumping laser for spin-electron generation. Our design integrates spintronic emitters with an optimized 1D \enquote{trapping} cavity, specifically engineered to fulfill critical aspects of ultrafast excitation and THz extraction. As a result, we have realized a 245\% enhancement in emission and an increase of 8~dB in overall intensity, positions our results among the most substantial improvements documented in this field. Furthermore, we delineate the optimal geometry for the deployment of STEs and explore the strategic selection of substrates in depth. Such enhanced emitters advance spintronic emitters towards broader applications in time-domain spectroscopy, ellipsometry, and nonlinear THz-pump spectroscopy. Enhancing spintronic emitter efficiency, combined with rapid magnetic field modulation, indicates the potential for dynamic ranges that rival traditional sources. Our predictions of STE's efficiencies, made through an electromagnetic approach, highlight its capability to uncover overlooked aspects from an optical standpoint, leading to subsequent improvements.

\section{Introduction}

Spintronic terahertz emitters (STE) are currently attracting considerable research attention due to their versatility, ease of fabrication, and cost-effectiveness \cite{seifert2016efficient,bull2021spintronic,wu2021principles,seifert2022spintronic}. These emitters operate based on a femtosecond optical excitation of a nanometric stack of typically magnetic and non-magnetic metallic films, leading to ultrafast diffusion of spins from a ferromagnetic (FM) layer into adjacent heavy nonmagnetic (NM) layers with strong spin-orbit coupling (SOC). The inverse spin-Hall effect (ISHE) deflection in the SOC NM layers is characterized by very fast relaxation times \cite{Kampfrath2013Sep} leading to a transient current dipole that is nearly Fourier-limited by the excitation pulse bandwidth. Femtosecond pump pulses thus generate near single cycle terahertz emission (Figure~\ref{fig:Schemes}a). ISHE STE's offer many advantages over other \enquote{traditional} pulsed THz sources, being unhindered by phonon absorption, phase-matching requirements, and long relaxation times. The groundbreaking work by Seifert \textit{et al.} demonstrated a broadband, gap-less spectrum spanning over 30 THz\cite{seifert2017ultrabroadband}.
ISHE THz pulses are independent of the pump wavelength and polarization, as they rely on an off-resonant generation of hot majority spin polarized carriers. This process depends only on the pump's fluence\cite{papaioannou2018efficient,herapath2019impact}. Moreover, these emitters are remarkably robust accommodating both collimated high-energy pulses on the order of millijoules at kHz repetition rates and tightly focussed low-energy pulses in the nanojoule range at sub-GHz repetition rates. At sub-MHz repetition rates, saturation and eventual thermal degradation sets in at a pump fluence of $\sim$5~mJ/cm$^2$\cite{vogel2022average}, while with increasing repetition rates the total average irradiance is the critical parameter. Highly confined fiber tip integrated STE's operate without degradation up to 1.5~kW/cm$^2$\cite{paries2023fiber}.   
These observations underline that ISHE sources possess a highly scalable flexibility allowing to use any pump laser system for its specific characteristics (repetition rate, wavelength, power, frequency tunability, pulse duration, etc\dots) \cite{yamahara2020ultrafast,takano2019terahertz}.  This is nicely illustrated by the latest record performance for W/CoFeB/Pt emitters. Exciting a large area emitter (5~cm diameter) with a collimated 2cm wide beam of 5~mJ pulses (35~fs, 800~nm, 1~kHz) from an amplified Ti:sapphire laser generates record focal THz peak electric fields exceeding 1.5~MV/cm in sub-ps cycles with an integrated average THz power close to 100~\textmu W\cite{Rouzegar2023}. Finally, with their easy lossless rotation of the emitted polarization plane by means of either a small rotating external magnet\cite{hibberd2019magnetic,kong2019broadband}, of a controlled Stoner-Wohlfarth rotation \cite{kolejak2022360}, or even piezoelectrically controlled magnetoelastic rotation \cite{lezier2022fully}, spintronic emitters have in a matter of just a few years taken an unique place as flexible, robust, ultrabroadband THz sources. Their polarization control has even allowed modulation bandwidths up to 10kHz and rates up to 10~MHz are within reach\cite{Gueckstock2021,Lezier2023May}.
The main drawback hindering a wider adoption of spintronic terahertz technology, is the low optical-to-terahertz efficiency, typically on the order of 10$^{-5}$-10$^{-6}$ with respect to the total input energy \cite{seifert2017ultrabroadband}. This is at least one to two orders of magnitude lower compared to traditional optically pumped terahertz sources \cite{fulop2020laser}. STE's invariably require fs-lasers operating at multiwatt average power levels. Enhancing their conversion efficiency is a critical step towards achieving competitiveness with conventional terahertz sources. 
In recent years, a lot of attention has been devoted to optimizing the pure spintronic aspects of the process: spin-current generation and spin-to-charge current conversion (S2C). Several excellent reviews give an overview of the different spin-generating and SOC materials that have been studied\cite{seifert2022spintronic,bull2021spintronic,Sinova2015Oct}. Up till now, it proves hard to radically improve upon the performance of the sturdy Pt(2nm)/CoFeB(1.8nm)/W(2nm)-system, though moderate increases have been reported by alloying the NM layers with Au to create spin sink effects and reduce spin reflection, and by replacing the FM spin injector by the semi-metal Heusler alloy Co$_2$MnGa for its very high degree of spin polarization\cite{hawecker2022spintronic}. A promising new direction is to move towards strong interfacial S2C that occurs by spin momentum locking in inversion symmetry-broken systems. Such Inverse Rashba-Edelstein-Effect (IREE) systems are currently under heavy investigation with reports on Bi$/$Ag interfaces\cite{jungfleisch2018control}, on surface effects of several topological insulators (SnBi$_2$Te$_4$\cite{Rongione2022}, Bi$_x$Sb$_{1-x}$\cite{Rongione2023}, and Bi$_2$Se$_3$\cite{tong2020enhanced}), and recently monolayer Pt-based transition metal dichalcogenides\cite{Abdukayumov2023}.
With IREE emitters still undergoing further optimization and the mature ISHE emitters being led by the Pt$/$CoFe$/$W family, further improvements must come from extrinsic parameters in the generation, i.e. the electromagnetic design that determines the absorption of the pulse energy and the radiation of the current dipole into free space. A bare trilayer emitter absorbs only about $30-50\%$ of the incident power in its spin-generation layer, and typical substrates are not impedance matched to free-space. Both factors combined reduce the generated charge current and the associated radiated THz E-field by nearly a factor $\sim 10$, which amounts up to two orders of magnitude in radiated THz power. Increasing the extracted THz radiation has been achieved by integrating a hemispherical substrate-matched collimating lens\cite{torosyan2018optimized} and by implementing antennas on the ISHE stack\cite{Nandi2019,Talara2021}, albeit at the cost of redistributing the THz spectrum. Recycling unabsorbed IR pump photons more efficiently is a logical strategy for enhancing performance. By introducing a photonic layered structure, the confinement of the optical pump field in the FM layer can be maximized with minimal impact on the emitted THz field. The incorporation of a dielectric Bragg-like mirror has led to reports of an efficiency increase of a single STE by a factor of 1.9 \cite{herapath2019impact,wagner2023optimised}. Cascading multiple STEs is another way of recycling pump photons though this comes at the cost of increased THz losses \cite{feng2018highly,jincascaded}. 
At present the research boom surrounding THz spintronics has led to a wealth of detailed insights concerning the intrinsic spintronic mechanisms, while the electromagnetic aspect of the devices' operation has remained less rigorously studied. Nevertheless, the optical perspective plays a crucial role in designing an optimal photonic geometry for the emitter. This applies both to the excitation and the emission processes at play. For instance, the pump absorption (and thus the efficiency) differs notably depending on whether the stack is excited from the substrate side or from the side of the metallic ISHE stack. Similarly, the emission is influenced by internal Fabry-Pérot reflections and scales differently towards either the substrate side or the air side. 
It is the goal of this work to present a rigorous electromagnetic theoretical framework for metallic W/CoFeB/Pt-based ISHE STE's and support this by experimental demonstrations of the predicted improvements. As the sub-picosecond dynamics of the emitter are a key aspect of its operation, we develop an original time-domain treatment correctly taking into account dispersion and non-stationary absorption. This theory is then applied for the design of an optimal structure that maximizes all aspects of the photonic processes at play. The experimental results confirm the numerical predictions by demonstrating THz peak field emissions enhanced by nearly 300\% with respect to a bare reference emitter.
The paper is organized as follows. We begin with presentation of our optimization strategy focused on electromagnetic efficiency. We delve into key processes such as excitation and emission in STEs, reconstruction of Fourier responses, and the comprehensive optimization of our STE-SiO$_2$/SiN-cavity samples. 
Further, we demonstrate the enhancement of our cavity-integrated spintronic emitters, comparing them to traditional STEs using time-domain spectroscopy and Fourier analysis. We also verify the linear correlation between THz emission and pump power in ISHE STE-Cavity emitters, and explore how excitation wavelength affects THz enhancement.
We demonstrate through simulation the critical role of the SiO$_2$ defect layer in cavity design, which is essential for phase alignment and interference control. We show the optimal thickness is key to ensuring the absorption spectrum aligns correctly with the pump pulse.
Thereafter, we delve into the efficiency of spintronic emitters, particularly in terms of THz extraction into free space and in the substrate direction. The performance and efficiency of spintronic terahertz emitters in various configurations, with and without photonic cavities, are thoroughly evaluated. The paper also examines how substrate choice impacts the efficiency of spintronic terahertz emitters, analyzing the effects of different substrates on absorption and THz emission. In conclusion, the paper presents a cohesive overview of the advancements and potential further improvements in the field of spintronic terahertz emitters.

\section{Results}

\subsection{Optimization strategy}

\begin{figure*}[!htb]
    \centering
\includegraphics[width=\textwidth]{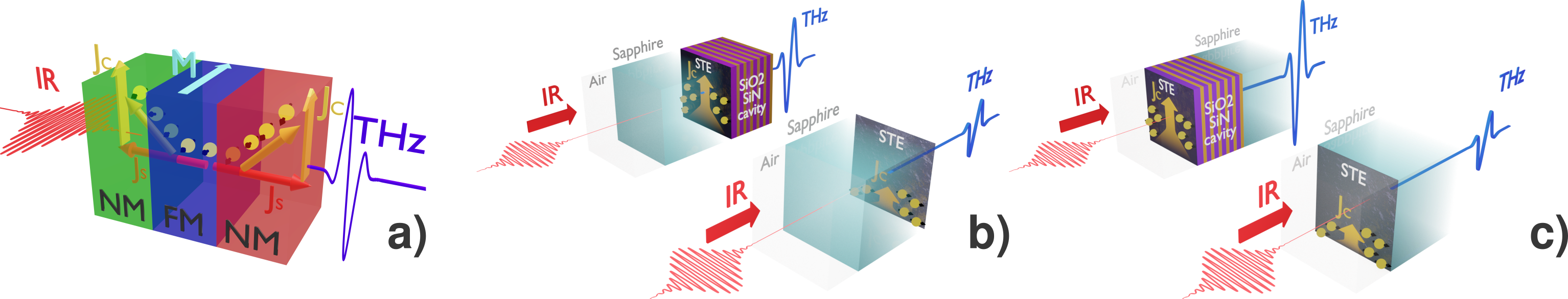}
    \caption{
a) ISHE Emitter Principle: IR pulse activates a NM/FM/NM tri-layer under magnetization $\mathbf{M}$, converting charge current $j_c$ to spin current $j_s$ and generating terahertz radiation. b) STE/Cavity/Sapphire Scheme: Illustrates the meaning of STE/Cavity/Sapphire sample excitation configuration in transmission geometry and its STE/Sapphire reference. c) Sapphire/STE/Cavity Scheme: Details the excitation arrangement of the Sapphire/STE/Cavity sample in transmission geometry with its Sapphire/STE reference.
    }
    \label{fig:Schemes}
\end{figure*}

The traditional STEs do not exploit their full potential and exhibit a rather limited absorption of only 30-50\% of the optical pump power. A significant portion of the incident power is wasted due to reflection and transmission losses. However, by effectively recycling the unused pump photons and optimizing the extraction of generated THz photons from the STE, we can achieve an extraordinary enhancement of the optical-to-terahertz efficiency. This electromagnetic perspective of STEs has never really been systematically studied. Their optimization has been primarily centered around spintronics, material engineering and magnetism \cite{seifert2022spintronic,feng2021spintronic,papaioannou2020thz,wu2021principles}. 
Therefore, we have developed a theoretical electromagnetic framework (described in detail in Supplemental Information), that rigorously treats the pulsed excitation and emission processes to optimize STEs. It is based on the following numerical steps: (I) Discretization of the given excitation spectrum into individual monochromatic plane waves (Fourier series); (II) Calculation of optical response of each individual Fourier term, employing the scattering matrix formalism; (III) Reconstruction of the Fourier responses to estimate the time-dependent spatial absorption through Poynting's theorem, providing valuable insight into the absorption dynamics; (IV) Utilization of this time-dependent absorption in the spin pumping layer as a source for temporal dipole emission by applying the scattering matrix formalism; (V) Global optimization of a) the IR pump absorption within the desired temporal interval and specific layers or of b) the output terahertz signal. The goal function (a) is well-suited for resolving problems related to single terahertz pulse emission, especially when increased absorption plays a direct role in the THz generation process. However, the optimization of the IR pump absorption can introduce detrimental side-effects on the THz extraction part of the electromagnetic problem, by for instance introducing extra THz losses, Fabry-Pérot effects or even multiple pulse effects. Therefore approach (b) is the more robust choice for optimizing an arbitrary structure to achieve maximum terahertz total power or peak power. It is however considerably more time and memory consuming. 
Using our approach an optimized photonic cavity is designed that efficiently traps nearly the entire spectral power of the pump laser in the magnetic spin pumping layer. For this trapping, we have opted for 1D SiO$_2$/SiN distributed dielectric layer pairs (a DBR mirror) as the resulting photonic crystal structure becomes only a few microns thick, impacting minimally the phase of the generated terahertz wavelengths and therefore minimally hindering terahertz extraction. The thicknesses of the individual layers of this photonic cavity were optimized for maximal enhancement using a global optimization (combination of Simplex and Differential evolution algorithms \cite{godlike,lagarias1998convergence}). Previous studies have primarily focused on augmenting absorption by placing a photonic \enquote{trapping} cavity on top of the STE and pumping the structure through the substrate. Such configurations have led to notable yet limited improvements of 60-80\% of the STE electric peak-to-peak field strength.\cite{herapath2019impact,jincascaded,wagner2023optimised}. However, our innovative approach takes into consideration an alternative excitation geometry, namely the \enquote{STE/(Cavity)/substrate arrangement}, implying a pump excitation from the metallic emitter side and a THz extraction through the substrate. Figures~\ref{fig:Schemes}b-c illustrate the difference in this new arrangement between the STE/Cavity/substrate design (a) and the common substrate/STE/Cavity (b), along with their corresponding references without cavities. It will be shown that this yields even higher levels of enhancement and superior overall performance. Notably, a recent study by Rouzegar et al. employed the STE/(Cavity)/Si configuration, leading to twofold improvement \cite{rouzegar2023broadband}. We will show that our approach achieves further emission improvements and also sheds light on many mechanisms from an optical perspective.
Specifically, we present a design that delivers an 2.45-fold terahertz field emission enhancement. This structure addresses the overlooked absorption disparity between STE/substrate (32\%) and substrate/STE (50\%). We also factor in the stronger THz field extraction into the substrate direction. In the next section, the experimental demonstration of this optimization strategy is presented.

\subsection{THz pulse peak-to-peak and power spectral density increase of PhC-enhanced STE's}~\\
\label{subsec:demonstration}
\begin{figure*}[!htb]
    \centering

\includegraphics[width=\textwidth]{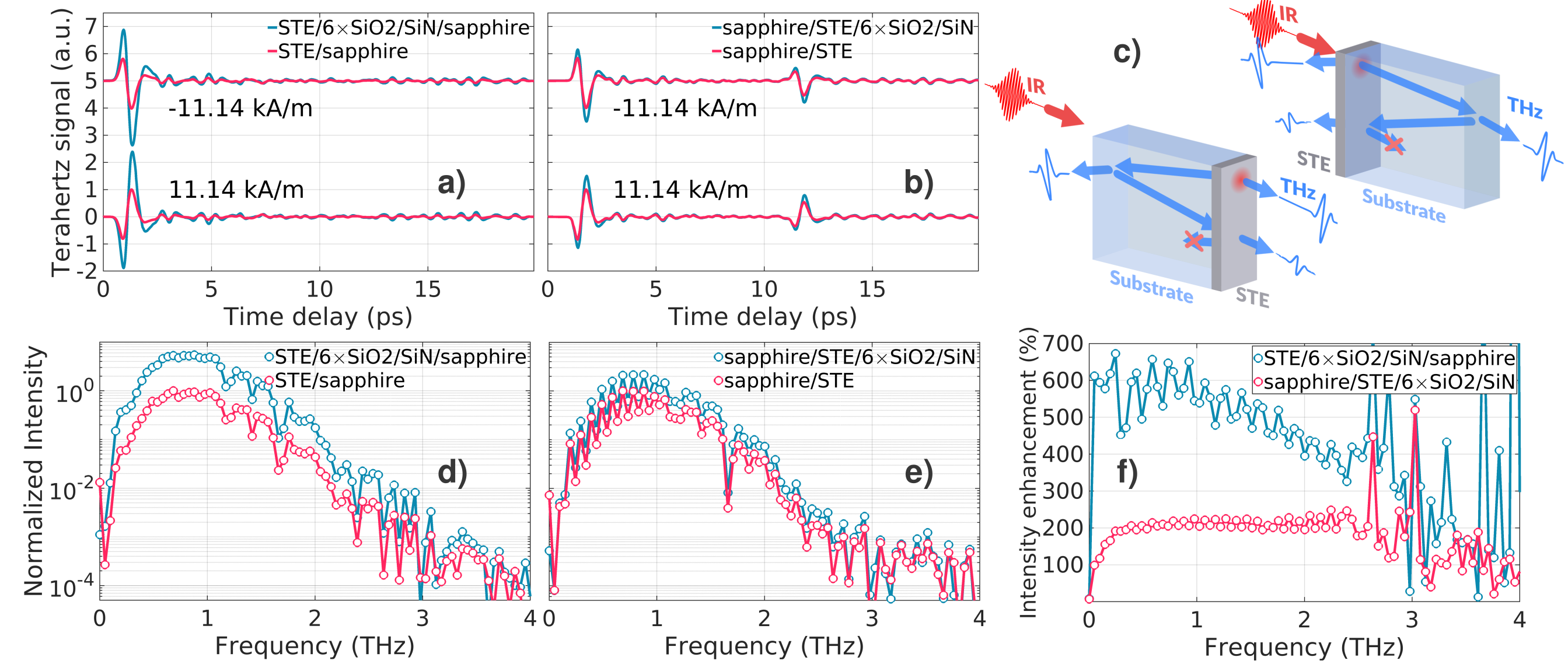}
    
    \caption{
a) Comparison of measured terahertz emission for STE/Cavity/sapphire sample and its STE/sapphire reference under magnetic fields of 11.14~kA/m and -11.14~kA/m. b) Measured terahertz emission comparison for sapphire/STE/Cavity sample and sapphire/STE reference in magnetic fields of 11.14~kA/m and -11.14~kA/m. c) Diagram explaining the origin of terahertz echoes in measured signals for sapphire/STE and STE/sapphire geometries. Where the anti-reflective coating effect suppresses direct reflections from the STE. d) Spectral intensity of the STE/Cavity/sapphire sample and comparison with its STE/sapphire reference. e) Spectral intensity comparison of the sapphire/STE/Cavity sample against its sapphire/STE reference. f) Shows the enhancement spectrum of STEs integrated with photonic cavities, comparing the STE/Cavity/sapphire and sapphire/STE/Cavity configurations. 
    }
    \label{fig:Measurements}
\end{figure*}
We performed thorough time-domain spectroscopy measurements to evaluate the emission performance and improvement of the sapphire(0.5mm)/STE/Cavity and STE/Cavity/sapphire(0.5mm) configurations in transmission geometry. The cavity consists of 6 SiO$_2$(139nm)/SiN(99.8nm) cells for 808nm, where the SiO$_2$ layer in the first cell is modified, and its design is explained in Section~\ref{subsec:design}. We utilize an optimized Pt(2.1 nm)/CoFeB(1.9 nm)/W(2 nm) trilayer as the STE emitter. More details on the experimental setup and the sample fabrication process are provided in the Supplementary Information. Figures~\ref{fig:Measurements}a-b plot the measured temporal traces for both geometries: (I) sapphire/STE/(Cavity) and (II) STE/(Cavity)/sapphire, pumped with 73~mW of average power. For both configurations, the THz temporal pulses generated by the reference emitter (non-augmented by a cavity), fabricated in the same run on the same substrate, are plotted in blue, $S_\mathrm{ref}(t)$. The red trace $S_\mathrm{PC}(t)$ represents the STE integrated with the optimized photonic cavity measured under the same pump power. Note that in these figures, the traces are normalized with respect to the peak amplitude of the reference emitter. In both cases the trapping photonic cavity clearly amplifies the peak-to-peak amplitude of the generated THz pulses, by a factor 1.5 for configuration (I) and a remarkable 2.45 for configuration (II). Previous attempts in the literature did not achieve such a significant increase in emission \cite{herapath2019impact,Rouzegar2023,feng2018highly}. As the spin-to-charge conversion depends in first instance on the energy delivered to the spin reservoir, the cost function for the design optimization is the electromagnetic absorption in the ferromagnetic layer. The optimization predicts absorbance increases from 50\% to 89\% for (I) (i.e. by $\times 1.78$) and from 32\% to 95\% for (II) ($\times 2.97$), respectively. The measured peak-to-peak enhancement $S_\mathrm{pp,PC} = A_\mathrm{pp} \times S_\mathrm{pp,ref}$  follows these predictions. Obviously, a full 100\% absorbance in the FM layer can not be reached because of unavoidable reflection losses and the impossibility to perfectly confine the pump field. We also provide evidence of the spintronic origin of the emitted terahertz radiation by observing a reversal in polarity when the magnetic field is reversed from 11.14~kA/m to -11.14~kA/m (vertically shifted traces in Figs~\ref{fig:Measurements}a-b).  Apart from the slight influence by water absorption, leading to the presence of minor long Lorentz oscillations in the time domain, a striking feature in these measurements is the appearance of a strong echo in the sapphire/STE/(Cavity) configuration (I). 
This reflected echo is suppressed in the STE/(Cavity)/sapphire configuration thanks to a broadband anti-reflection phenomenon \cite{kroll2007metallic,ding2016ultrathin,carli1977reflectivity}, as schematically shown in Figure~\ref{fig:Measurements}c. Forward-emitted THz pulses, when reflecting off the bottom of the substrate, produce weaker replicas that are delayed by twice the substrate's propagation time ($\Delta\tau = 2n_{\mathrm{Al}2\mathrm{O}3}d{\mathrm{Al}_2\mathrm{O}_3}/c\approx 10$~ps). However, when pulses reflect off thin metallic layers, they result in almost perfect antiphase interference rather than a simple reflection. This effect is attributed to the nanometer thickness of these layers and the imaginary component of their Drude THz permittivity.
The squared modulus of the Fourier transforms of the TDS signals presented in Figs.\ref{fig:Measurements}d and e, reconfirm clear evidence of signal improvement across the entire spectrum of the STE signal, while the noise level remains unchanged. Not only is there a significant increase of the dynamic range in configuration (II), its spectrum is devoid from interference fringes. Contrary to an electrooptic sampling, exact knowledge of the calibrated transfer function of the THz photoconductive receiver is not straightforward to determine\cite{Kampfrath_EOS_TF}. As a result, absolute values for the transient electric field strength of the THz pulse are hard to extract out of our measurements. However, we can estimate the relative power increase of the PhC enhanced STE's by integrating the ratio of the TDS spectra $|\mathcal{S}_\mathrm{PC}(\omega)|^2/|\mathcal{S}_\mathrm{ref}(\omega)|^2$ (Figs\ref{fig:Measurements}d-e). 
These integrated ratios yield values of 2.12 and 5.64, which closely match the increase predicted by merely considering the square of the peak-to-peak signal enhancement $A_{pp}^2 = 1.5^2=2.25 \text{ and } 2.45^2=6.0025 $. The frequency-dependent ratios before integration illustrated in Fig.~\ref{fig:Measurements}f, suggest a consistent enhancement across the spectrum. This is attributed to the fact that the proportional change in absorption do not impact the dynamics of ISHE emission, and thus does not influence the THz spectrum. This phenomenon is observed in the Sapphire/STE/Cavity sample, but not in the STE/Cavity/Sapphire sample, for reasons that remain unclear to us. One possible explanation might be the different sequences in the fabrication of the STE and the cavity. In the STE/Cavity/sapphire configuration, the cavity is exposed to higher temperatures during STE deposition, potentially leading to changes in material properties. At low THz frequencies, the enhancement actually reaches predicted 600\% compared to the integrated average of 564\%, suggesting a decrease in efficiency at higher frequencies. However, at these higher frequencies, the signal is markedly weaker, which makes the inconsistent efficiency reduction almost unnoticeable in the time domain.
Configuration (II) increases the power conversion efficiency of Pt/CoFeB/W-emitters by over 8~dB, marking a 50\% improvement on the record by Rouzegar et al \cite{rouzegar2023broadband}. Our operation at 80~mW and 80~MHz results in pulses with a maximum energy of 2.0~nJ after the chopper, limiting demonstration to TDS temporal trace analysis. Yet, extrapolating Rouzegar's findings suggests our configuration could potentially achieve $>$150\textmu W average THz power with high-energy pumping. This would make spintronic THz emitters competitive with, and superior in pulse bandwidth to, conventional pulsed THz sources.

\subsection{Pump power and wavelength dependence}~\\
\begin{figure*}[!th]
    \centering
    \includegraphics[width=\textwidth]{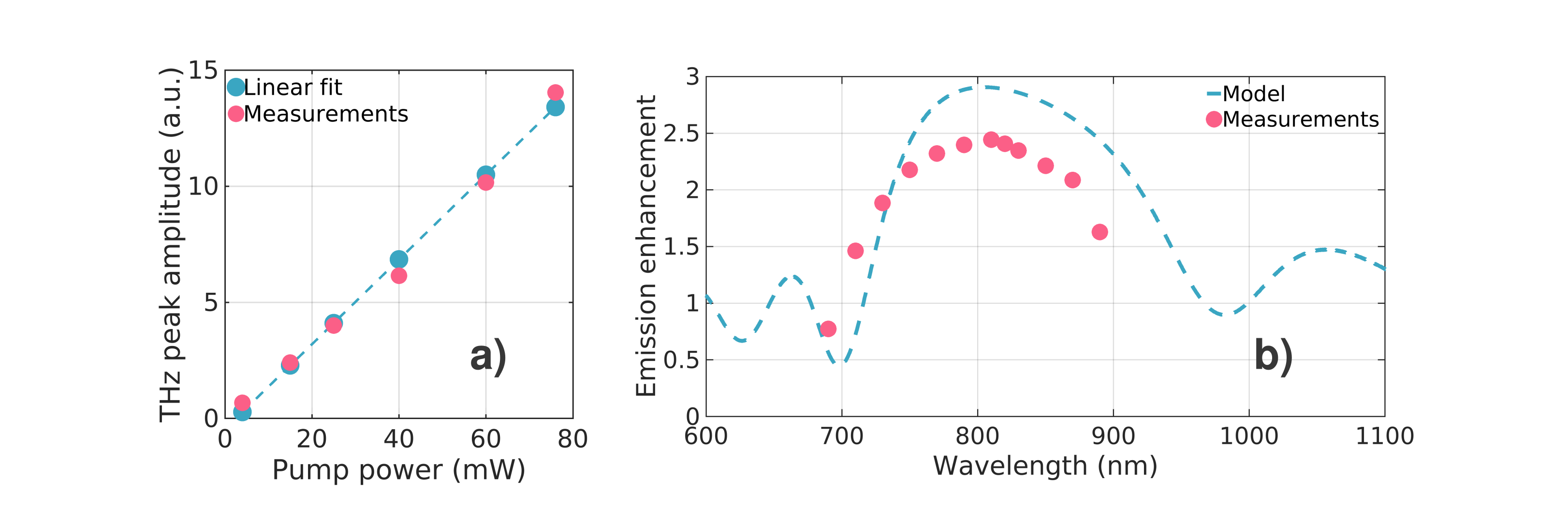}
    \caption{
Demonstration of the robustness of STE/Cavity/sapphire emitter: a) Generated electric field's linear response to varying average powers of an 808~nm, 100~fs IR pump pulse. b) Enhancement of terahertz amplitude with changing central wavelengths of the excitation pulse (under the same power) compared to model expectations.  
    }
    \label{fig:Sweeps}
\end{figure*}
The transient current created in ISHE emitter is intrinsically proportional to the absorbed energy in the spin pumping layer. The THz field should thus be linearly dependent on the pump power. To validate this, we conducted an experimental power sweep using a tunable attenuator. This power adjustment was performed behind the chopper (see Supplemental Information). The power was swept from 0 to 80~mW. Fig.\ref{fig:Sweeps}a plots the variation of the peak-to-peak TDS signal as a function of pump power (measured separately before each experiment). Over the entire range, a perfect linear dependence is observed, confirming the functional efficiency of cavities the structural integrity and non-saturating behaviour with increasing power. This is unsurprising as the pumping conditions ($\sim$0.05~mJ/cm$^2$ maximal fluence) are very conservative, compared to the reported damage thresholds\cite{vogel2022average,paries2023fiber}.
An important advantage of the inverse spin Hall effect is its independence with respect to the center wavelength of the pump pulse \cite{papaioannou2018efficient}. This is a direct consequence of the superdiffusive spin transport models that have successfully described the THz transient dipole current in ISHE STE's \cite{Battiato2010Jul}. The amplitude of the spin current at its origin is primarily determined by the energy dumped in the electron bath of the FM layer which has a broad largely wavelength independent absorption spectrum (see Supplemental Information). The wavelength dependence of the STE is therefore dominated by the electromagnetic response of the cavity system to the pump. This is again a clear argument in favour of devoting more attention to this aspect. 
We performed emission analysis while adjusting the excitation wavelength of the tunable Ti:sapphire oscillator Mai Tai-HP to investigate its impact on the predicted THz enhancement. The oscillator's pulse duration is guaranteed to be sufficiently constant over its tuning range. It is therefore expected that the THz peak-to-peak enhancement will emulate the absorption spectrum of the designed photonic structure. This is presented in Fig.~\ref{fig:Sweeps}b, plotting the ratio of the peak THz values of the STE/Cavity/sapphire and STE/sapphire samples as a function of central wavelength. In doing so, we factor out the wavelength dependence of the average power of the oscillator. The measured enhancement follows qualitatively the modelled absorption spectrum of the designed photonic cavity and especially reproduces correctly the predicted spectral position of the gap of the DBR mirror. Even though the model predicts a 20\% higher enhancement, the observation of maximum efficiency at 808~nm suggests accurate fabrication of the structure's bandgap and a reliable model used for the layers optimization strategy. 
Thermal effects might explain the discrepancy between theoretical models and actual measurements. The ISHE's conversion efficiency is very low, around $10^{-5} - 10^{-6}$, meaning almost all pump power must be dissipated. Although sapphire substrates in traditional ISHE STEs serve as decent heat sinks, the thermal resistance significantly increases with the addition of a SiO$_2$/SiN DBR mirror \cite{Burghartz1994Sep,Shan2003Jun,Talghader2004Apr,Sinha1978Apr,Kumar2017Jun}. This, coupled with a threefold increase in power absorption, likely impacts the device's thermal management, an aspect not yet considered in our analysis but warrants further investigation.
\section{Discussion}
Having experimentally demonstrated the 250\% THz field enhancement induced by the photonic cavity in the STE/cavity/substrate configuration, we delve a bit deeper into the design considerations having led to this optimised structure. This discussion will equally reveal why both pumping configurations (STE/cavity/substrate versus substrate/STE/cavity) behave drastically different despite nearly equal pump absorption. This section will conclude by an outlook on how to further improve STE emitters from an electromagnetic point of view.%
\subsection{Integrating PhC with STEs: design and optimization}
\label{subsec:design}
\begin{figure*}[!th]
    \centering
    \includegraphics[width=\textwidth]{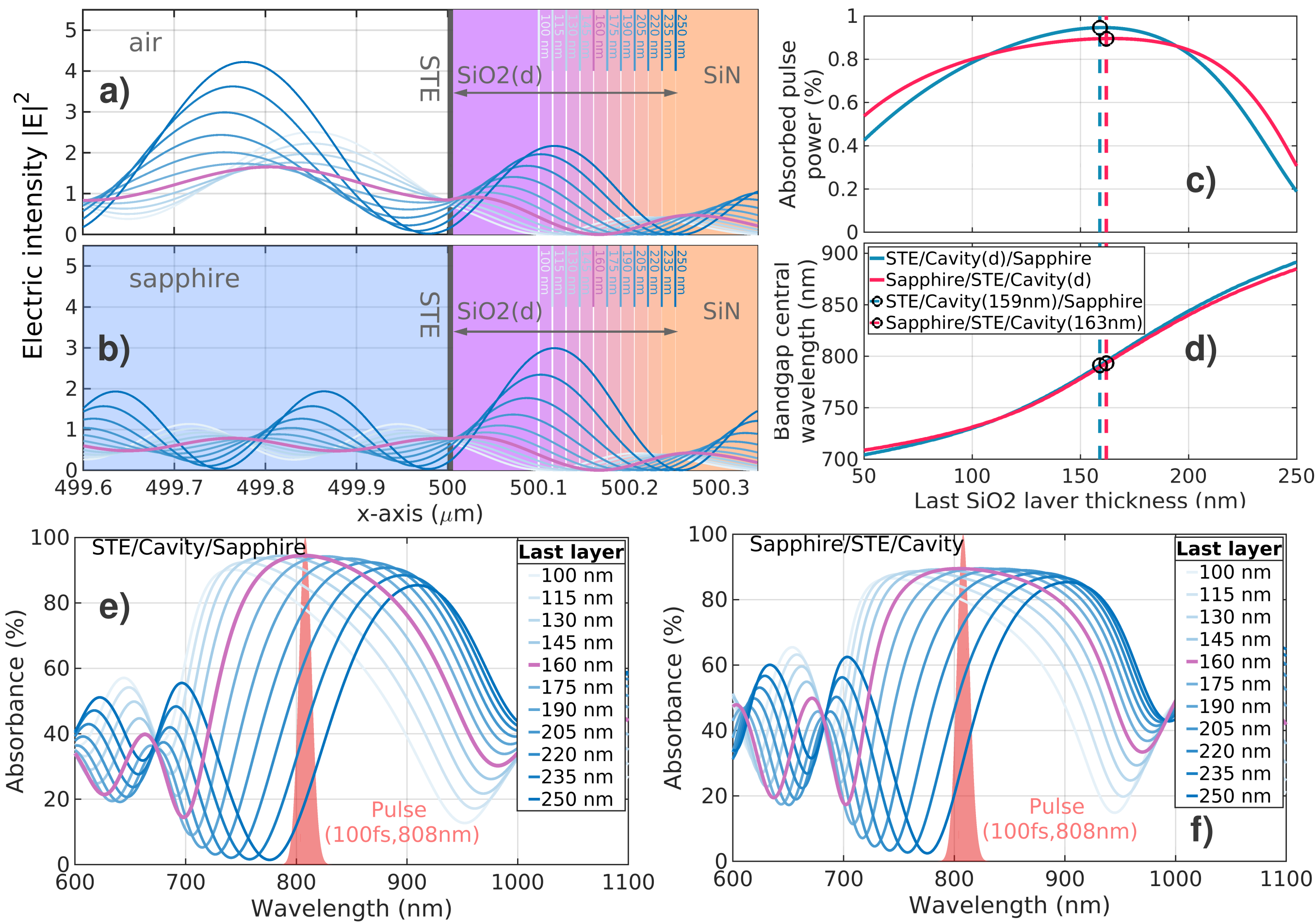}
    \caption{
Sensitivity analysis of STE/Cavity/sapphire and sapphire/STE/Cavity to SiO$_2$ layer thickness:
a-b) Visualization of the field distribution transformation (blue shades) near the spintronic layers, showing changes in SiO$_2$ layer thickness from 100-250~nm. The optimal configuration (160~nm - magenta) demonstrates maximum field intensity within the spintronic layers for both configurations.
c) Comparison of changes in the absorbed energy of an excitation pulse (100~fs, 808~nm) and (d) the central absorption spectrum wavelength with different SiO$_2$ layer thicknesses. Optimized solutions are marked.
e-f) Absorption spectrum transformation in STE/Cavity/sapphire and sapphire/STE/Cavity structures with varying SiO$_2$ thickness. The shaded blue curves illustrate the bandgap shift across wavelengths, while the magenta line approximates the 160~nm solution corresponding to a pulse (100~fs, 808~nm), whose spectrum is shown for comparison.
    }
    \label{fig:LastLayer}
\end{figure*}

Our cavity design features a SiO$_2$/SiN DBR mirror, enhanced by an additional SiO$_2$ layer acting as a defect layer. This layer is crucial as it introduces an extra variable to counteract the phase shift from the Bragg mirror, thus fine-tuning the interference within the FM layer. Our experiments show that six pairs of SiO$_2$/SiN are adequate for effective trapping, with additional pairs not further improving FM layer absorption. The thickness of the SiO$_2$ last layer was optimized to 159~nm for the sapphire/STE/Cavity and 162~nm for the STE/Cavity/sapphire, resulting in configurations of (I): Sapphire/\allowbreak STE/\allowbreak SiO$_2$(23nm)/\allowbreak 6$\times$[SiO$_2$(139.0nm)/SiN(99.8nm)] and (II): STE/\allowbreak SiO$_2$(20.0nm)/\allowbreak 6$\times$[SiO$_2$(139.0nm)/SiN(99.8nm)]/\allowbreak Sapphire.
In both configurations (I) and (II), only the layers of the (Cavity) structure were varied during the optimization procedure. The Bragg mirror converges to a 239~nm thick unit cell: SiO$_2$(139.0~nm)/SiN(99.8~nm). This is a logical consequence of optimizing a trapping structure centered around 808~nm. Each layer in the unit cell presents approximately a $\lambda/4$ thickness at 808~nm \cite{malitson1965interspecimen,luke2015broadband}.
 To demonstrate the impact of the last SiO$_2$ layer's thickness, we conducted simulations over a range of 100 to 250~nm. This layer plays a crucial function in compensating for the phase change induced by metallic spintronic layers within the entire photonic cavity, applicable to both STE/Cavity/Sapphire and Sapphire/STE/Cavity architectures. The results of this fine-tuning compensation are evident in the alteration of the infrared field distribution, as shown in Fig.~\ref{fig:LastLayer}a-b, with varying the SiO$_2$ layer thickness. When the thickness is not optimally set, the interference maximum is positioned outside the spintronic layers, leading to significant field reflection and passage through the cavity, and consequently, low overall absorption. While the SiO$_2$/SiN cavity is engineered to produce a bandgap in reflection at 808~nm,  the misalignment in phase across different STE/defect configurations results in the absorption bandgap shifting to different wavelengths. This shift occurs when the standing wave at these wavelengths fulfills the criteria for enhancing the field in the spintronic emitter, but not necessarily for a pulse at 808~nm.
 Figure~\ref{fig:LastLayer}e-f illustrates the transformation of the absorption spectrum (shown in shades of blue) compared to the spectrum of a 100~fs pulse with a central wavelength of 808~nm. After optimizing the SiO$_2$ layer, approximately 160~nm for both cases, including a 20~nm defect layer, the absorption spectrum is aligned with the pulse spectrum. However, at the optimal thickness the interference maximum of the field is aligned within the spintronic layers in Fig.~\ref{fig:LastLayer}a-b. This alignment effectively minimizes reflection and transmission from the structure. 
In a more direct approach, we demonstrate the changes in total absorption and the corresponding alterations in the central wavelength of the absorption spectrum on the varying thickness of the SiO$_2$ layer. Figure~\ref{fig:LastLayer}c-d effectively illustrates these changes, where each marked point corresponds to an optimized solution for a 100~fs excitation pulse. The optimized solutions we obtained correspond to the peaks in the absorption curves. It is crucial to emphasize that the thickness dependency is not overly acute, thereby enabling fabrication within a few nanometers precision. As anticipated, the solution's bandgap central wavelength is found to be around 800~nm. 
The design and fabrication of structures for sub-100~fs pulses is inherently more challenging. Shorter pulses exhibit a broader spectrum to align with an equally broad bandgap. For pulses as short as about 10~fs (approximating 3 micrometers), their dimensions become comparable to the cavity size itself. In these scenarios, the temporal evolution of interference patterns, especially during the reflection of the pulse from the cavity, emerges as a critical aspect. Hence, the design of cavities must prioritize femtosecond pulse considerations over the central wavelength of the pulse.
 Note that this distinguishes this design from a Tamm plasmon surface state between a metal and a DBR mirror\cite{KaliteevskiPRB2007,Jiang2023}. A Tamm plasmon resonance searches to create a surface eigenstate with a moderate quality factor confined near a metal but at in-plane wave vectors accessible by direct optical excitation. Here the goal is to maximize the absorption inside the metal of the incidence pulse rather than to achieve a resonant field buildup near the metal. It can be seen from Fig.~\ref{fig:LastLayer}e-f  that the pump absorption is not peaked at any wavelength but high over the entire pump pulse spectrum. 


\subsection{Generation process}

\label{subsec:generation}
\begin{figure*}[!th]
    \centering
\includegraphics[width=0.95\textwidth]{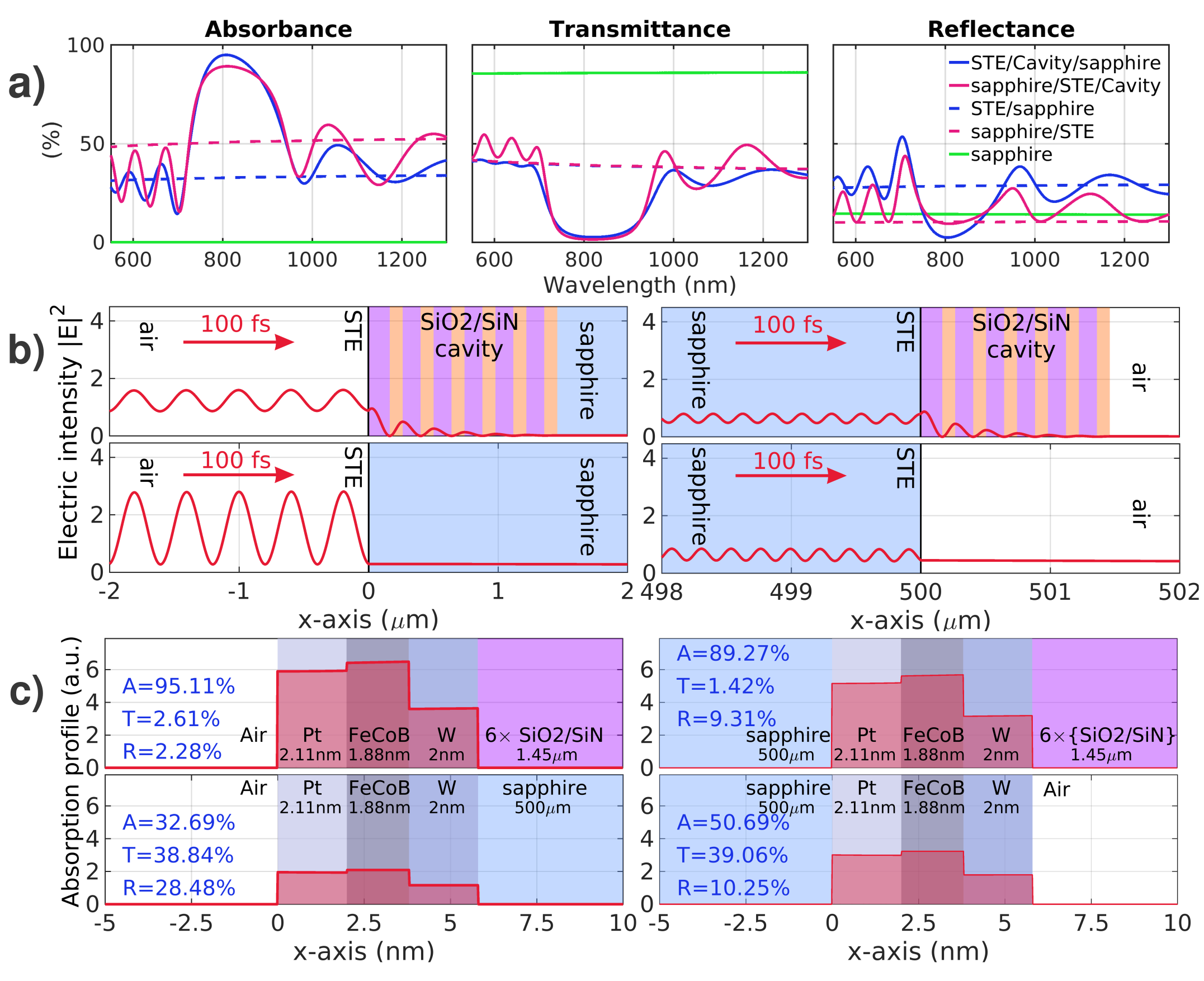}

    \caption{
Analysis of absorption during IR excitation:
a) Calculated absorbance, transmittance, and reflectance from extracted material properties for the studied sample configurations, shown in a non-coherent approximation (excluding coherent interference with substrate reflection).
b) Distribution of the $|E|^2$ field during excitation with a 100-fs pulse (arrow indicates direction) in STE-Cavity and their STE references, demonstrating changes in the reflected, absorbed, and transmitted intensities and field distribution within spintronic layers.
c) Absorption distribution $Q(x,t)$ across Pt, CoFeB, W layers. Each subplot's left side summarizes the excitation pulse conversion into absorption, transmission, and reflection for each configuration.
    }
    \label{fig:Generation}
\end{figure*}
To provide further insights into the performance of both STE geometries, we present the estimated spectral absorbance, transmittance, and reflectance in Fig.~\ref{fig:Generation}a using the measured optical constants (see Supplemental Information), specifically comparing a conventional sapphire-based STE (dashed lines) with a cavity-enhanced STE (continuous lines).  In Fig.~\ref{fig:Generation}a the red and blue curves refer to respectively pumping from the substrate or the spintronic stack. Due to reciprocity the transmittance of a bare emitter is independent of its excitation sense. One would therefore be tempted to conclude that the excitation sense with the lowest reflectance, i.e. sapphire/STE, will inevitably lead to the strongest THz emission. As will be seen in Sec.\ref{subsec:extraction}, this is not necessarily true. The integration of a cavity in both excitation geometries creates a bandgap that nearly perfectly recycles the pump photons: the transmittance through the DBR mirror is in both cases lower than 3\% over the bandgap. For a comprehensive overview, we present the integrated total absorbance ($A$), transmittance ($T$), and reflectance ($R$) of the IR pulse (100~fs, 808~nm) in Fig.~\ref{fig:Generation}b (the pair on the right), accounting for contributions from Fabry-Pérot reflections and the effects of all interfaces.
In the substrate/STE/cavity configuration, the DBR cavity effectively reduces the transmission of pump photons and converts it to absorption, but it does not significantly reduce the internal reflection from the STE layers, which are inherently low.
The total reflectance of the structure remains at least always at the same level as without the cavity (comparing the full and dashed red curves in Fig.~\ref{fig:Generation}a). In this scenario, the main factor contributing to reflectance is the reflection at the air/sapphire interface, which accounts for losses of $\approx$7.6\%. Conversely, the reflection from the STE itself makes a minimal contribution, amounting to just a few percent, even in the absence of a cavity. This is fundamental consequence of the different phase of the first Fresnel reflection at the entrance interface of the STE stack. This is a first reason why exciting the STE through substrate is less optimal. From an electromagnetic point of view, even by adding a trapping cavity, the pump photons will always undergo a minimal \enquote{entrance} reflection. For the STE/cavity/substrate configuration on the other hand, the total reflectance is drastically reduced within the DBR gap, yet it still comprises a few percent. This fundamentally different behaviour limits the maximal achievable absorption when exciting a STE through an optically dense layer.
Time-domain simulations confirm the above frequency domain findings. Figures~\ref{fig:Generation}b plot snapshots of the intensity distribution $|E|^2$ (red line) halted at the instant of peak intensity inside the spin pumping layer upon excitation by a 100~fs pulse (30~\textmu m long, centred at 808~nm). The red arrow denotes the direction of excitation. The interference of the exciting and back-reflected fields appears nicely, revealing for instance how in the substrate/STE/cavity configuration the interference ripple is \emph{not} reduced. The cavity does manage to recycle the pump photons that would otherwise be transmitted, but it can't do anything about the IR photons that escape back towards to the sapphire at the STE/sapphire interface. Hence the absence of pump reflection reduction in this configuration (as already shown in Fig.~\ref{fig:Generation}a).
Additionally, our time-domain framework allows to plot the spatial distribution of the energy absorption as a function of time. Fig.~\ref{fig:Generation}c shows a zoom of the structure on the metallic stack, plotting on top of it the absorption distribution  $Q(x,t)$, paused at the maximal field fluence in the FM layer (see Supplemental Information for the definition and the physical meaning of $Q$). 
This simulated $Q(x,t)$ snapshot shows that even though our structure manages to radically improve the absorption in the spin pumping layer (see the twofold and threefold $Q$ increase in the CoFeB layer in Fig.~\ref{fig:Generation}c), the optimization strategy cannot avoid that an important part of the absorbed pump energy is lost in the non-magnetic non-spin polarized layers. In detail, the energy is distributed among the Pt/CoFeB/W layers in the following ratio: 37.0\% for Pt, 40.3\% for CoFeB, and 22.7\% for W. This is a fundamental limitation due to the optical properties of Pt and W. Remember that a variation of the thicknesses of these layers might heavily impact the efficiency of the inverse Spin Hall effect itself. As Pt and W provide the spin-orbit coupling for the spin-to-charge conversion their thickness is determined by their spin diffusion length \cite{Tao2018Jun,Yu2018Jul}.

\subsection{THz extraction/emission}
\label{subsec:extraction}
\begin{figure*}[!th]
    \centering
   \includegraphics[width=\textwidth]{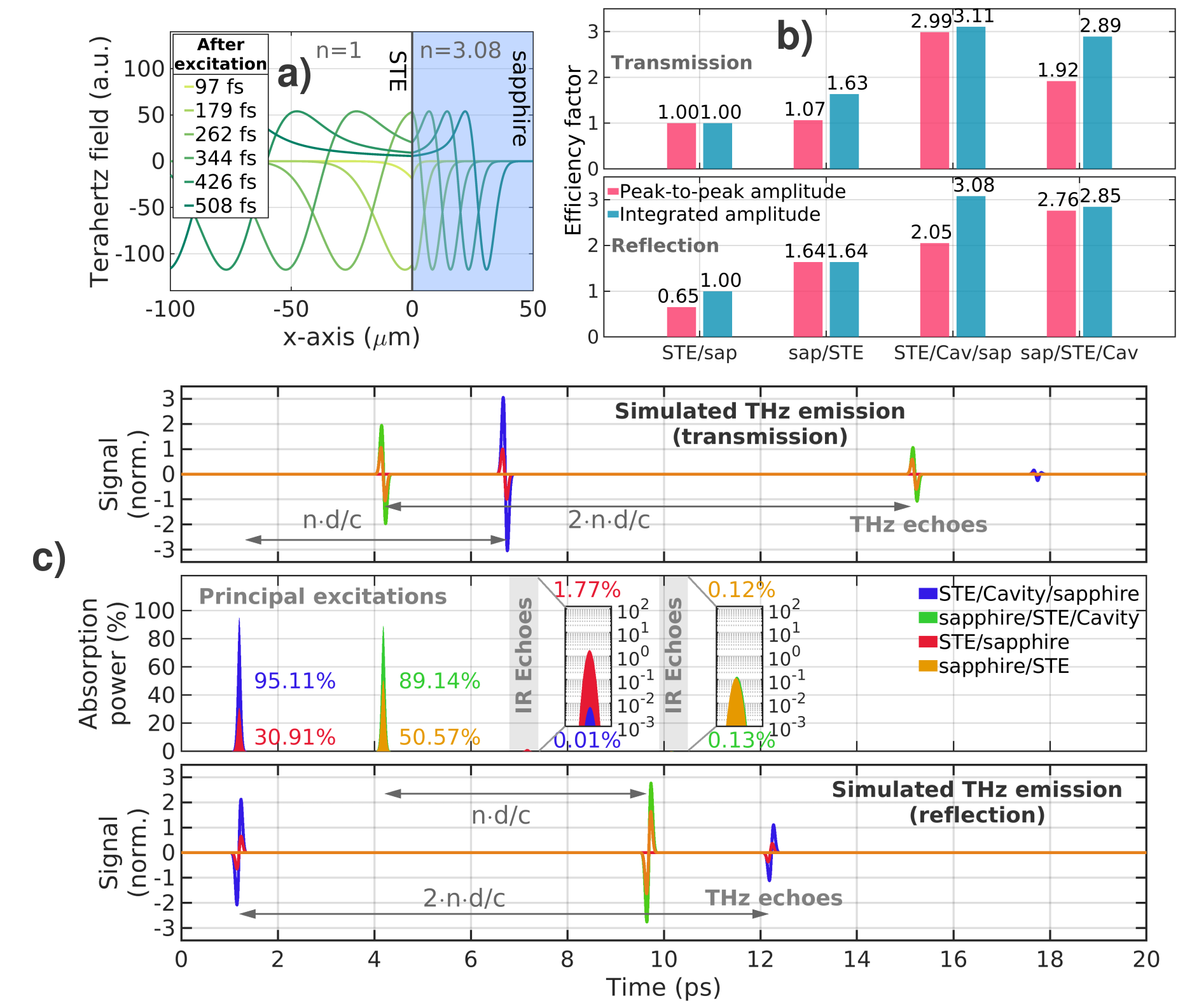} 
    \caption{
a) Simulation illustrates THz field development (shades of green) emitted from spintronic emitter layers into sapphire and air, displayed at various moments post-IR pump excitation (97-508~fs), showing consistent amplitude in both mediums and differing intensities based on $I=\frac{1}{2}cn\epsilon_0 \mathbf{E}\cdot \mathbf{E}^*$.
b) Middle subplot: shows dynamic absorption over time for all samples from Poynting's theorem, with the Y-axis scaled to reflect absorption percentage. The absorpted power is indicated by the area under each curve. IR echoes are marginal and detailed in logarithmic scale. Top and bottom subplots: depict calculated THz emissions from these absorptions, directed in transmission and reflection, respectively, all normalized against the STE/Sapphire emission.
c) Comparison of efficiency for different samples in reflective and transmissive configurations, comparing peak-to-peak values in the time domain as well as integrated full wavefront area in the time domain.
    }
    \label{fig:Extraction}
\end{figure*}
So far, our focus has been primarily on optimizing pump absorption, specifically in terms of Terahertz (THz) generation, without discussing the electromagnetic details related to the emission of THz into free space. Now, our investigation shifts to a detailed analysis of the true efficiency of spintronic emitters including the extraction.
The conventional spintronic emitter demonstrates comparable peak-to-peak effectiveness, comparing principal pulses in both the STE/sapphire and sapphire/STE geometries, as experimentally demonstrated by the transmission measurements in Figure~\ref{fig:Measurements}a. This observation might appear paradoxical, considering the significant discrepancies in pump absorption across the different geometries (32\% versus 50\%). Hence, we would logically expect the primary pulse to be stronger in the sapphire/STE configuration when a cavity is absent. However, a radiating dipole between two different media, oriented parallel to their interface, will have a continuous electric (tangential) field independent of the refractive index of the medium into which it radiates. To illustrate this concept, we conducted simulations of the dipole emission and field distribution throughout the radiation process emanating from spintronic layers situated between sapphire and air. The procedure for conducting the THz emission simulation will be explained in the subsequent section. Figure~\ref{fig:Extraction}a plots the terahertz pulse fields at multiple intervals post-excitation by an infrared pulse. In both sapphire and air, the fields reach the expected similar magnitudes. It's illustratively evident that the field in sapphire is scaled due to its higher permittivity.
Therefore, even though the electric fields reach the same values in both the sapphire and air media, the electromagnetic intensity $\mathcal{E}$ is scaled by the refractive index $n$ 
 \begin{equation}
    \mathcal{E}=\frac{1}{2}c \epsilon_0 n \, \mathbf{E} \cdot \mathbf{E}^*, 
 \end{equation}
where $c$ represents the velocity of light, $\epsilon_0$ is the vacuum permittivity and $\mathbf{E}$ denotes the complex electric amplitude. In a simple approximation, a pumped STE emits an intensity into its surrounding media in proportion to the respective conductances of these media. If the substrate is semi-infinite, the ratio of the radiated intensities between substrate and air superstrate is thus $Z_\mathrm{vac}/Z_\mathrm{sub}$, where $Z=Z_0/n$ is the impedance \cite{Saleh1991Aug}. However, terminating the substrate with an air interface reduces this emission by a Fresnel power transmission coefficient, given by $4Z_\mathrm{vac}Z_\mathrm{sub}/(Z_\mathrm{vac} + Z_\mathrm{sub})^2$ under normal incidence \cite{Button1979}. For conventional STE's, the absorption ratio is approximately $0.31/0.5$ between STE/sapphire and sapphire/STE, resulting in a radiated intensity ratio of $\left(\frac{0.31}{0.5}\right)^2=0.39$. In this idealized scenario, this ratio is almost compensated by the intensity ratio (represented by $4Z_\mathrm{vac}^2/(Z_\mathrm{vac}+Z_\mathrm{sub})^2$), yielding $2.25 \times 0.39 = 0.87$.


\subsection{STE vs. STE-Cavity: efficiency in different geometries}
%
%
In our effort to assess the performance of spintronic terahertz emitters both with and without cavities in transmission and reflection geometries, we focus on the dynamics of absorption under femtosecond excitation and its distribution over time to reconstruct the emission profile, leading to the efficiencies. This approach is detailed in Supplemental Information. So far, our representation of absorption has primarily been through frequency-dependent spectra and visual snapshots of field distributions, as demonstrated in Fig.~\ref{fig:Generation}a-c. Now, we calculated the time-dependent absorption power $P_{ab}(t)$ for all studied configurations: STE/sapphire vs. sapphire/STE, and STE/Cavity/sapphire vs. sapphire/STE/Cavity, demonstrated in Fig.~\ref{fig:Extraction}c (middle subplot). The percentage of absorbed energy is indicated by the marked areas under the curves, with the vertical axis scaled to approximate the percentage of absorbed power, correlating directly with amplitude. 
The absorption profiles in the time-domain, along with the previous analysis of absorption spectra in Fig.\ref{fig:Generation}a, indicate that most absorption occurs within the principal excitation. Although we do observe Fabry-Pérot reflections, their contribution is typically less than 2\%, as detailed upon in the logarithmic scale.  This results in only a marginal fraction of energy reaching the STE for Fabry-Pérot absorption and secondary emission. The time delay in the principal excitations of the spintronic layers between the STE/sapphire and sapphire/STE configurations is a consequence of the excitation pulse propagating through the 500~µm-thick sapphire substrate. Finally, the radiating dipole field of spintronic emitters is determined by taking the derivative of the dipole source, as $E_\mathrm{dipole}\propto \frac{\partial P_{ab}}{\partial t}$.
Subsequently, the emitted field propagates through the specified structures using Scattering matrices, as detailed in the Supplemental Information. Furthermore, we can numerically predict emission for both reflection and transmission, even though our experimental setup is limited to transmission measurements. All simulation results for reflection (displayed in the bottom subplot) and transmission (in the top subplot) are  presented in Fig.\ref{fig:Extraction}c, normalized concerning the highest observed peak-to-peak value in the STE/sapphire configuration for transmission. This normalization serves to enable a realistic comparison of which configurations are effective relative to each other, rather than merely discussing improvements over the configuration without a cavity. The adoption of time-dependent absorption guarantees the consideration of absorption scaling and the treatment of all THz and IR reflections (echoes, losses), leading to valuable estimations of simulated efficiencies. As an illustration, the simulations of the conventional STE/sapphire and sapphire/STE configurations in Fig.~\ref{fig:Extraction}c exhibit identical behaviour in the transmission arrangement as experimental results in Fig.~\ref{fig:Measurements}a-b. Their normalized peak-to-peak efficiency in the time domain are comparable a close to one. 
Evaluating signals based on their maximal peak-to-peak in the time domain is crucial for generating intense pulses in spectroscopy to obtain a wide dynamic range, and equally important for nonlinear applications that require an intense pump. However, it becomes evident that the sapphire/STE configuration includes an echo, contributing to the overall energy and resulting in a larger energy yield in this configuration. The origin of this echo was explained in Sec.~\ref{subsec:demonstration} and in the diagram Fig.~\ref{fig:Measurements}c. The contribution of the echo alters the efficiencies in all integrated pulses to 1.63 and extends the possibilities for applications in the spectral domain, particularly amplifying at resonant frequencies corresponding to the substrate thickness. Therefore, we will present a summary of both the normalized peak-to-peak and integrated waveform efficiency in Fig.~\ref{fig:Extraction}b.
When we redirect the emission of conventional emitters towards a reflective direction, it becomes evident that the balance between absorption and THz efficiency is disrupted. In such scenario, the STE/sapphire configuration exhibits relatively low absorption (31\%) and decreased THz emission into the surrounding air (refer to Sec~\ref{subsec:generation}), resulting in the lowest efficiency 0.65 (1.00 for all integrated pulses) when compared to other configurations. Conversely, the sapphire/STE arrangement stands out as the most effective basic configuration of 1.64 efficiency due to its significant absorption (50\%) and intense THz extraction towards the substrate, leading to the integrated efficiency of 1.64. These simulations indicate that for scenarios excluding a cavity, the sapphire/STE configuration in reflective geometry offers the most promise. Moreover, it provides echo-free time domains, which are particularly advantageous for data processing in THz-TDS spectroscopy or THz-pump experiments, where a secondary THz-pump could interfere with the ongoing process.
For STE configurations incorporating cavities, the STE/Cavity/sapphire configuration in a transmission geometry emerges as the superior choice. It showcases the highest absorption at 95\% and elevates emission towards the substrate, resulting in an efficiency of 2.99 (and 3.11 when considering integrated waveform). This is followed by sapphire/STE/Cavity in reflection, where the cavity limits absorption to no more than 89\%, but the radiation is directed into the sapphire, achieving an efficiency of 2.76 (2.85 in integrated waveforms). The efficiencies are the lowest and comparable in the configurations of sapphire/STE/Cavity in transmission (1.92 and 2.89) and for STE/Cavity/sapphire in reflection with efficiencies (2.05 and 3.08).
\begin{figure*}[h!]
    \centering
    \includegraphics[width=\textwidth]{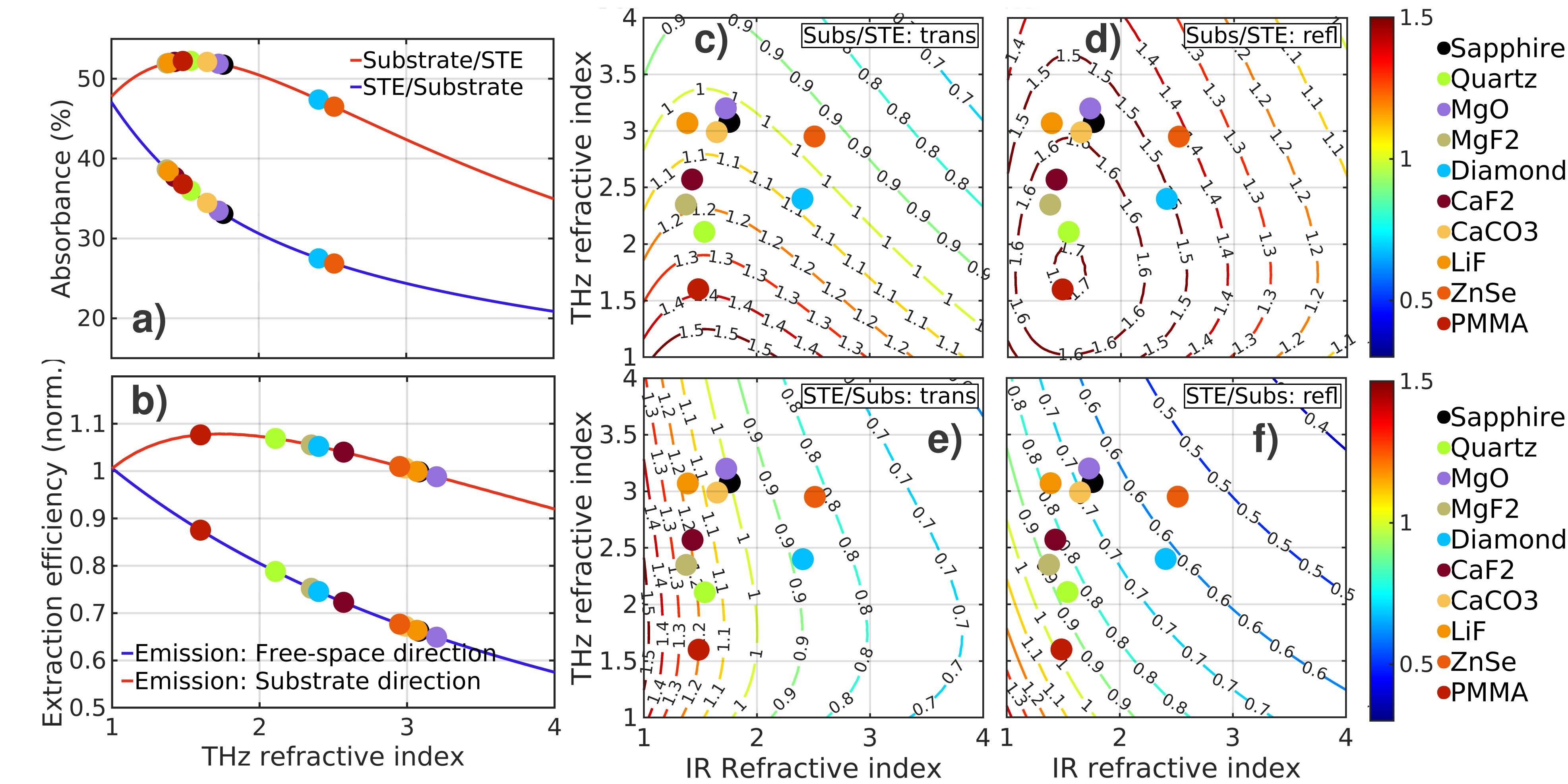}

    \caption{ Efficiency impact of substrate selection in Spintronic Emitters: The graphs illustrate efficiency's dependence on the refractive index within the THz/IR range. Each simulation compares typical substrate materials to illustrate their performance. Efficiencies are normalized against the STE/sapphire transmission configuration for straightforward comparison. 
    a) Simulated absorbance change of a 100~fs, 808~nm pulse in Substrate/STE and STE/Substrate configurations. 
    b) Simulated efficiency of dipole emissions from spintronic layers, both into free space and into the substrate. c-f) Efficiency maps of configurations:  
    c) Substrate/STE in transmission, 
    d) Substrate/STE in reflection, 
    e) STE/Substrate in transmission, 
    f) STE/Substrate in reflection.
    }
    \label{fig:Substrates}
\end{figure*}

\subsection{Further improvements and outlook}
In the final section, we investigate how selecting the optimal substrate can further enhance the performance of spintronic emitters, leading to variations in both excitation and extraction efficiency. The selection of a substrate influences the pump's absorption in thin spintronic layers that are directly deposited onto it.  Figure~\ref{fig:Substrates}a illustrates how absorbance in the spintronic layers of a standard emitter changes with the refractive index of the substrate at 800nm. While the STE/substrate configuration exhibits a decreasing trend in absorbance as the refractive index increases, due to greater reflection from the optically dense material of the metal films, the substrate/STE configuration reaches a peak absorption of 50\% at a refractive index around 1.7. This maximum results from the combined effect of enhanced transmission through the optically denser substrate via STE and the reflection loss at the air/substrate interface.
The simulation compares several commonly used substrate materials  \cite{Marple1964Mar,MGO,Ghosh1999May,Li1980Jan,Phillip1964Nov,sapphire,Tsuda2018Mar}, which are listed in the legend. Sapphire is highly suitable for the Substrate/STE configuration, as are materials like quartz, MgO, MgF$_2$, and others. In the STE/Substrate configuration, materials with lower refractive indices, like MgF$_2$ or LiF, are associated with increased absorption. However, Substrate/STE consistently results in higher absorptions. In the substrate/STE configuration, efficiency is reduced as the IR pulse gets reflected at the air/Substrate interface, which results in a loss described by $|(Z_\mathrm{vac}-Z_\mathrm{sub})/(Z_\mathrm{vac}+Z_\mathrm{sub})|^2$. To mitigate this effect, an anti-reflective coating can increase efficiency by an additional 3-20\% for the mentioned substrates.
Subsequently, the choice of substrate also impacts the extraction of terahertz emission from spintronic layers in two significant ways. Firstly, (I) THz emission is scaled by a factor of $1/(n_\mathrm{vac}+n_\mathrm{sub}+Z_0\int_0^d\sigma(z)\mathrm{d}z)$ \cite{seifert2016efficient}. This implies that an increase in the substrate's refractive index leads to a decrease in the emitter's overall efficiency. (II) As previously discussed in Sec.~\ref{subsec:extraction}, the emission towards the substrate is $Z_\mathrm{vac}/Z_\mathrm{sub}$-fold more intense compared to emission into the air.
In the free-space direction, THz emission always decreases with an increasing refractive index. Conversely, towards the substrate, emission increases (point II) until the overall reduction in emission outweighs this benefit (point I). This trend is evident in the simulated peak-to-peak efficiency in the time-domain, as shown in Fig.~\ref{fig:Substrates}b. To better assess relative enhancements, we normalize our results against the widely-used STE/sapphire transmission configuration (towards substrate), allowing for a clearer comparison. Geometry selection is key and both STE/substrate (reflection) and substrate/STE (transmission) geometries face notable limitations in achieving optimal terahertz efficiency. Yet the evident effect of substrate choice on THz efficiency cannot be overlooked. 
In the simulation involving typical substrates \cite{Rogalin2018Dec,Sajadi2015Nov,Tcypkin2020Oct,Grischkowsky1990Oct,Palik1997,Fan2015Oct}, materials with low refractive indices like PMMA show the highest performance.  Quartz and MgF$_2$ are also promising as substrate materials. Conversely, sapphire, when evaluated for THz efficiency, proves to be a less suitable option. In addition, reflective losses at the interface between substrate and air are notably higher for the listed materials, in the range of 3-30\%. This is linked to their increased refractive index in the THz range, which can be partially balanced by using an anti-reflection coating designed for THz frequencies.
To better understand the overall peak-to-peak efficiency of common spintronic emitters, which depends on both IR absorption and THz extraction, we present contour maps in Fig.~\ref{fig:Substrates}c-f illustrating STE efficiency based on THz-IR refractive indices. For comparative clarity, efficiencies across all geometries are normalized against the STE/sapphire configuration in transmission geometry. The map includes various common materials, enabling a comparison of the effects of different substrates.
The Substrate/STE configuration in reflection emerges as the most efficient, showing up to a 1.7-fold enhancement. This peak efficiency, achieved through effective extraction and absorption, aligns with refractive indices near 1.5. Such indices are typical for materials like plastics, demonstrated by PMMA, and extend to others such as Teflon, TPX, Zeonex, and Topas \cite{Du2021Nov,Islam2020May}.  In this context, enhancing efficiency through terahertz extraction is more advantageous than focusing on absorption, which reaches its maximum potential with cavity integration. Nonetheless, the STE/Substrate configuration still significantly improves absorption. The least efficient is the STE/Substrate configuration in reflection, where both extraction and absorption are ineffective, requiring materials with very low refractive indices.
Clearly, the appropriate selection of materials is key to further enhancing efficiency. However, increasing the refractive index beyond 1.5-2 typically leads to a decrease in efficiency. Our simulations didn't account for the absorption coefficient, which can greatly affect efficiency, particularly at higher THz frequencies. Attention is needed for reflective losses at the substrate/air interface, which can be mitigate for both THz and IR spectra with the use of anti-reflective coatings. Apart from these materials' optical efficiency, factors like surface quality and adhesion properties are crucial, impacting the deposition and integrity of spintronic layers \cite{nenno2019modification}. Moreover, the importance of thermal conductivity for passive cooling and the damage threshold cannot be overstated, with sapphire being a practical choice for these properties \cite{Shan2003Jun,Burghartz1994Sep}. The choice of substrate also dictates the fabrication process and, consequently, the cost.

\section{Conclusion }
This research was dedicated to the improvement in the performance of Pt/CoFeB/W spintronic emitters, with the objective of maximizing their efficiency to its fullest potential. 
These emitters are becoming more acknowledged within the terahertz community for their versatile applications, cost-effectiveness, and user-friendliness. A significant drawback of these emitters, compared to traditional sources, is their optical-to-terahertz conversion efficiency, which is lower by several orders of magnitude. However, due to the ultrafast demagnetization process, these emitters often achieve peak performance levels that are exceptionally high. Despite high peak performance, the significant efficiency limitations of these spintronic emitters hinder their extensive use in spectroscopy, nonlinear applications, and communications, limiting the full utilization of their potential.
Therefore, our objective was to maximize the enhancement of spintronic emitters by integrating them with 1D SiO$_2$/SiN photonic cavities, aiming to enable these sources to rival conventional ones. We developed a robust and comprehensive methodology for optimizing these devices, thereby introducing a new perspective to address this technical challenge. The design of our cavities played a pivotal role in significantly enhancing the trapping of excitation photons. This enhancement was crucial in amplifying the generation of spin electrons, thereby intensifying the spintronic emission. Our fabricated samples demonstrated an exceptional absorption rate of approximately 95\% of the excitation pulse and a substantial enhancement of the emitter's THz amplitude, achieving 2.45 times the performance of traditional spintronic emitters, along with an 8 dB increase in energy output – a distinct achievement in this field. We provide a comprehensive guide detailing the design of these structures, alongside sharing our developed solutions of STE-Cavities, which are finely tuned for optimal performance. Additionally, through our analysis of the fabricated samples, which involved varying the pump power and adjusting the central wavelength of the excitation pulse, we demonstrated the robustness and operational effectiveness of these enhanced spintronic emitters.
This study is the first to provide a detailed comparison of the STE/sapphire/Cavity with the sapphire/STE/Cavity configuration. Our theoretical approach aids in explicating the key factors that influence emitter efficiency and electromagnetic functionality. We offer a thorough evaluation of emitter efficiencies in both reflective and transmission geometries, essential for researchers in selecting the most suitable emitter configuration, given the substantial differences in efficiency. Our findings reveal, for example, that the pure emitter's maximum efficiency, without a cavity, is significantly higher for substrate/STE configuration in reflection – showing nearly 60\% more efficiency compared to its transmission geometry. Furthermore, our results establish that superior performance is exclusively attainable with the STE/Cavity/sapphire configuration, which effectively combines maximal absorption and enhanced terahertz (THz) emission directed towards the substrate.
Despite the promising findings, it's crucial to acknowledge that our access was limited to an 80 MHz oscillator, restricting our ability to test the spintronic emitters under high-energy millijoule (mJ) excitation pulses. This limitation leaves potential high-energy impacts, like saturation and thermal damage, unexplored. Furthermore, we acknowledge that our simulations might overestimate the results compared to empirical data, possibly overlooking aspects such as heat generation, saturation effects, or the sensitivity to the materials used in the simulations. Therefore, a crucial point was characterizing the spintronic layers' properties, which we conducted through ellipsometry and transmission measurements. Despite these limitations, we are confident that our simulations provide a realistic representation of the spintronic emitters' behavior.
Finally, we analyzed the impact of substrates on spintronic emitters and propose ways to further enhance their efficiency. Our simulation strategy presents a promising perspective, not limited to just augmenting excitation via increased absorption, but also extends to the careful optimization of structures for optimal terahertz field extraction, incorporating techniques like anti-reflective coatings and terahertz cavities.
In closing, this study marks a substantial advancement in optimizing spintronic emitters for terahertz applications.  Our results also provide a practical guide to understanding spintronic emitters from the optical standpoint. The specific findings on cavity design, efficiency analysis, substrate influence, and pulse-based simulations give rise to new possibilities for enhancing emitter efficiency and functionality, setting a clear direction for future developments in this field.
\section*{Funding}
The authors acknowledge financial support from the Horizon 2020 Framework Programme of the European Commission under FET-Open Grant No. 863155 (s-Nebula). The infrastructure used was made available through project No. CZ.02.01.01/00/22\_008/0004631- "Materials and Technologies for Sustainable Development," funded by the European Union and the state budget of the Czech Republic within the framework of the Jan Amos Komensky Operational Program. The research was supported by Czech Science Foundation Grant No. 22-33060S). The research was supported by Czech Science Foundation Grant No. 22-33060S). This work was supported by the Ministry of Education, Youth and Sports of the Czech Republic through the e-INFRA CZ (ID:90254). 

\section*{Acknowledgement}
The authors thank the RENATECH Network for the support in the realization of the devices.

\bibliographystyle{ieeetr} 
\bibliography{bibliography.bib}

\begin{thebibliography}{10}

\bibitem{seifert2016efficient}
T.~Seifert, S.~Jaiswal, U.~Martens, J.~Hannegan, L.~Braun, P.~Maldonado, F.~Freimuth, A.~Kronenberg, J.~Henrizi, I.~Radu, {\em et~al.}, ``Efficient metallic spintronic emitters of ultrabroadband terahertz radiation,'' {\em Nature photonics}, vol.~10, no.~7, pp.~483--488, 2016.

\bibitem{bull2021spintronic}
C.~Bull, S.~M. Hewett, R.~Ji, C.-H. Lin, T.~Thomson, D.~M. Graham, and P.~W. Nutter, ``Spintronic terahertz emitters: Status and prospects from a materials perspective,'' {\em APL Materials}, vol.~9, no.~9, p.~090701, 2021.

\bibitem{wu2021principles}
W.~Wu, C.~Yaw~Ameyaw, M.~F. Doty, and M.~B. Jungfleisch, ``{Principles of spintronic THz emitters},'' {\em Journal of Applied Physics}, vol.~130, no.~9, p.~091101, 2021.

\bibitem{seifert2022spintronic}
T.~S. Seifert, L.~Cheng, Z.~Wei, T.~Kampfrath, and J.~Qi, ``Spintronic sources of ultrashort terahertz electromagnetic pulses,'' 2022.

\bibitem{Kampfrath2013Sep}
T.~Kampfrath, K.~Tanaka, and K.~A. Nelson, ``{Resonant and nonresonant control over matter and light by intense terahertz transients},'' {\em Nat. Photonics}, vol.~7, pp.~680--690, Sept. 2013.

\bibitem{seifert2017ultrabroadband}
T.~Seifert, S.~Jaiswal, M.~Sajadi, G.~Jakob, S.~Winnerl, M.~Wolf, M.~Kl{\"a}ui, and T.~Kampfrath, ``{Ultrabroadband single-cycle terahertz pulses with peak fields of 300 kV cm$^{-1}$ from a metallic spintronic emitter},'' {\em Applied Physics Letters}, vol.~110, no.~25, p.~252402, 2017.

\bibitem{papaioannou2018efficient}
E.~T. Papaioannou, G.~Torosyan, S.~Keller, L.~Scheuer, M.~Battiato, V.~K. Mag-Usara, J.~L’huillier, M.~Tani, and R.~Beigang, ``{Efficient terahertz generation using Fe/Pt spintronic emitters pumped at different wavelengths},'' {\em IEEE Transactions on Magnetics}, vol.~54, no.~11, pp.~1--5, 2018.

\bibitem{herapath2019impact}
R.~I. Herapath, S.~M. Hornett, T.~Seifert, G.~Jakob, M.~Kl{\"a}ui, J.~Bertolotti, T.~Kampfrath, and E.~Hendry, ``Impact of pump wavelength on terahertz emission of a cavity-enhanced spintronic trilayer,'' {\em Applied Physics Letters}, vol.~114, no.~4, p.~041107, 2019.

\bibitem{vogel2022average}
T.~Vogel, A.~Omar, S.~Mansourzadeh, F.~Wulf, N.~M. Saban{\'e}s, M.~M{\"u}ller, T.~S. Seifert, A.~Weigel, G.~Jakob, M.~Kl{\"a}ui, {\em et~al.}, ``{Average power scaling of THz spintronic emitters efficiently cooled in reflection geometry},'' {\em Optics Express}, vol.~30, no.~12, pp.~20451--20468, 2022.

\bibitem{paries2023fiber}
F.~Paries, N.~Tiercelin, G.~Lezier, M.~Vanwolleghem, F.~Selz, M.-A. Syskaki, F.~Kammerbauer, G.~Jakob, M.~Jourdan, M.~Kl\"{a}ui, Z.~Kaspar, T.~Kampfrath, T.~S. Seifert, G.~von Freymann, and D.~Molter, ``Fiber-tip spintronic terahertz emitters,'' {\em Opt. Express}, vol.~31, pp.~30884--30893, Sep 2023.

\bibitem{yamahara2020ultrafast}
K.~Yamahara, A.~Mannan, I.~Kawayama, H.~Nakanishi, and M.~Tonouchi, ``Ultrafast spatiotemporal photocarrier dynamics near gan surfaces studied by terahertz emission spectroscopy,'' {\em Scientific reports}, vol.~10, no.~1, p.~14633, 2020.

\bibitem{takano2019terahertz}
K.~Takano, M.~Asai, K.~Kato, H.~Komiyama, A.~Yamaguchi, T.~Iyoda, Y.~Tadokoro, M.~Nakajima, and M.~I. Bakunov, ``Terahertz emission from gold nanorods irradiated by ultrashort laser pulses of different wavelengths,'' {\em Scientific Reports}, vol.~9, no.~1, p.~3280, 2019.

\bibitem{Rouzegar2023}
R.~Rouzegar, A.~Chekhov, Y.~Behovits, B.~Serrano, M.~Syskaki, C.~Lambert, D.~Engel, U.~Martens, M.~Münzenberg, M.~Wolf, G.~Jakob, M.~Kläui, T.~Seifert, and T.~Kampfrath, ``{Broadband Spintronic Terahertz Source with Peak Electric Fields Exceeding 1.5 MV/cm},'' {\em Physical Review Applied}, vol.~19, p.~034018, 3 2023.

\bibitem{hibberd2019magnetic}
M.~Hibberd, D.~Lake, N.~Johansson, T.~Thomson, S.~Jamison, and D.~Graham, ``Magnetic-field tailoring of the terahertz polarization emitted from a spintronic source,'' {\em Applied Physics Letters}, vol.~114, no.~3, p.~031101, 2019.

\bibitem{kong2019broadband}
D.~Kong, X.~Wu, B.~Wang, T.~Nie, M.~Xiao, C.~Pandey, Y.~Gao, L.~Wen, W.~Zhao, C.~Ruan, {\em et~al.}, ``Broadband spintronic terahertz emitter with magnetic-field manipulated polarizations,'' {\em Advanced Optical Materials}, vol.~7, no.~20, p.~1900487, 2019.

\bibitem{kolejak2022360}
P.~Kolej{\'a}k, G.~Lezier, K.~Postava, J.-F. Lampin, N.~Tiercelin, and M.~Vanwolleghem, ``{360$^\circ$ polarization control of terahertz spintronic emitters using uniaxial FeCo/TbCo$_2$/FeCo trilayers},'' {\em ACS Photonics}, vol.~9, no.~4, pp.~1274--1285, 2022.

\bibitem{lezier2022fully}
G.~Lezier, P.~Kolej{\'a}k, J.-F. Lampin, K.~Postava, M.~Vanwolleghem, and N.~Tiercelin, ``{Fully reversible magnetoelectric voltage controlled THz polarization rotation in magnetostrictive spintronic emitters on PMN-PT},'' {\em Applied Physics Letters}, vol.~120, no.~15, p.~152404, 2022.

\bibitem{Gueckstock2021}
O.~Gueckstock, L.~Nádvorník, T.~S. Seifert, M.~Borchert, G.~Jakob, G.~Schmidt, G.~Woltersdorf, M.~Kläui, M.~Wolf, and T.~Kampfrath, ``{Modulating the polarization of broadband terahertz pulses from a spintronic emitter at rates up to 10 kHz},'' {\em Optica}, vol.~8, p.~1013, 7 2021.

\bibitem{Lezier2023May}
G.~Lezier, P.~Kolej{\ifmmode\acute{a}\else\'{a}\fi}k, P.~Kolej{\ifmmode\acute{a}\else\'{a}\fi}k, J.-F. Lampin, K.~Postava, M.~Vanwolleghem, and N.~Tiercelin, ``{Ultrafast Modulation of Polarization in Spintronic THz Emitters Enhanced by Field Induced Spin Reorientation Transition},'' {\em Optica Publishing Group}, p.~SF2I.4, May 2023.

\bibitem{fulop2020laser}
J.~A. F{\"u}l{\"o}p, S.~Tzortzakis, and T.~Kampfrath, ``Laser-driven strong-field terahertz sources,'' {\em Advanced Optical Materials}, vol.~8, no.~3, p.~1900681, 2020.

\bibitem{Sinova2015Oct}
J.~Sinova, S.~O. Valenzuela, J.~Wunderlich, C.~H. Back, and T.~Jungwirth, ``{Spin Hall effects},'' {\em Rev. Mod. Phys.}, vol.~87, pp.~1213--1260, Oct. 2015.

\bibitem{hawecker2022spintronic}
J.~Hawecker, E.~Rongione, A.~Markou, S.~Krishnia, F.~Godel, S.~Collin, R.~Lebrun, J.~Tignon, J.~Mangeney, T.~Boulier, {\em et~al.}, ``{Spintronic THz emitters based on transition metals and semi-metals/Pt multilayers},'' {\em Applied Physics Letters}, vol.~120, no.~12, p.~122406, 2022.

\bibitem{jungfleisch2018control}
M.~B. Jungfleisch, Q.~Zhang, W.~Zhang, J.~E. Pearson, R.~D. Schaller, H.~Wen, and A.~Hoffmann, ``{Control of terahertz emission by ultrafast spin-charge current conversion at Rashba interfaces},'' {\em Physical review letters}, vol.~120, no.~20, p.~207207, 2018.

\bibitem{Rongione2022}
E.~Rongione, S.~Fragkos, L.~Baringthon, J.~Hawecker, E.~Xenogiannopoulou, P.~Tsipas, C.~Song, M.~Mičica, J.~Mangeney, J.~Tignon, T.~Boulier, N.~Reyren, R.~Lebrun, J.~M. George, P.~L. Fèvre, S.~Dhillon, A.~Dimoulas, and H.~Jaffrès, ``{Ultrafast Spin-Charge Conversion at SnBi$_2$Te$_4$/Co Topological Insulator Interfaces Probed by Terahertz Emission Spectroscopy},'' {\em Advanced Optical Materials}, vol.~10, p.~2102061, 4 2022.

\bibitem{Rongione2023}
E.~Rongione, L.~Baringthon, D.~She, G.~Patriarche, R.~Lebrun, A.~Lema{\ifmmode\hat{\imath}\else\^{\i}\fi}tre, M.~Morassi, N.~Reyren, M.~Mi{\ifmmode\check{c}\else\v{c}\fi}ica, J.~Mangeney, J.~Tignon, F.~Bertran, S.~Dhillon, P.~Le~F{\ifmmode\acute{e}\else\'{e}\fi}vre, H.~Jaffr{\ifmmode\grave{e}\else\`{e}\fi}s, and J.-M. George, ``{Spin-Momentum Locking and Ultrafast Spin-Charge Conversion in Ultrathin Epitaxial Bi$_{1-x}$Sb$_x$ Topological Insulator},'' {\em Adv. Sci.}, vol.~10, p.~2301124, July 2023.

\bibitem{tong2020enhanced}
M.~Tong, Y.~Hu, Z.~Wang, T.~Zhou, X.~Xie, X.~Cheng, and T.~Jiang, ``{Enhanced terahertz radiation by efficient spin-to-charge conversion in Rashba-mediated Dirac surface states},'' {\em Nano Letters}, vol.~21, no.~1, pp.~60--67, 2020.

\bibitem{Abdukayumov2023}
K.~Abdukayumov, M.~Mičica, F.~Ibrahim, L.~Vojáček, C.~Vergnaud, A.~Marty, J.-Y. Veuillen, P.~Mallet, I.~G. de~Moraes, D.~Dosenovic, S.~Gambarelli, V.~Maurel, A.~Wright, J.~Tignon, J.~Mangeney, A.~Ouerghi, V.~Renard, F.~Mesple, J.~Li, F.~Bonell, H.~Okuno, M.~Chshiev, J.-M. George, H.~Jaffrès, S.~Dhillon, and M.~Jamet, ``{Atomic-Layer Controlled Transition from Inverse Rashba–Edelstein Effect to Inverse Spin Hall Effect in 2D PtSe$_2$ Probed by THz Spintronic Emission},'' {\em Advanced Materials}, p.~2304243, 2024.

\bibitem{torosyan2018optimized}
G.~Torosyan, S.~Keller, L.~Scheuer, R.~Beigang, and E.~T. Papaioannou, ``{Optimized spintronic terahertz emitters based on epitaxial grown Fe/Pt layer structures},'' {\em Scientific reports}, vol.~8, no.~1, pp.~1--9, 2018.

\bibitem{Nandi2019}
U.~Nandi, M.~S. Abdelaziz, S.~Jaiswal, G.~Jakob, O.~Gueckstock, S.~M. Rouzegar, T.~S. Seifert, M.~Kläui, T.~Kampfrath, and S.~Preu, ``Antenna-coupled spintronic terahertz emitters driven by a 1550 nm femtosecond laser oscillator,'' {\em Applied Physics Letters}, vol.~115, 2019.

\bibitem{Talara2021}
M.~Talara, D.~S. Bulgarevich, C.~Tachioka, V.~K. Mag-Usara, J.~Muldera, T.~Furuya, H.~Kitahara, M.~C. Escaño, Q.~Guo, M.~Nakajima, G.~Torosyan, R.~Beigang, M.~Watanabe, and M.~Tani, ``{Efficient terahertz wave generation of diabolo-shaped Fe/Pt spintronic antennas driven by a 780 nm pump beam},'' {\em Applied Physics Express}, vol.~14, p.~042008, 3 2021.

\bibitem{wagner2023optimised}
F.~M. Wagner, S.~Melnikas, J.~Cramer, D.~A. Damry, C.~Q. Xia, K.~Peng, G.~Jakob, M.~Kl{\"a}ui, S.~Ki{\v{c}}as, and M.~B. Johnston, ``{Optimised Spintronic Emitters of Terahertz Radiation for Time-Domain Spectroscopy},'' {\em Journal of Infrared, Millimeter, and Terahertz Waves}, pp.~1--14, 2023.

\bibitem{feng2018highly}
Z.~Feng, R.~Yu, Y.~Zhou, H.~Lu, W.~Tan, H.~Deng, Q.~Liu, Z.~Zhai, L.~Zhu, J.~Cai, {\em et~al.}, ``Highly efficient spintronic terahertz emitter enabled by metal--dielectric photonic crystal,'' {\em Advanced Optical Materials}, vol.~6, no.~23, p.~1800965, 2018.

\bibitem{jincascaded}
Z.~Jin, Y.~Peng, Y.~Ni, G.~Wu, B.~Ji, X.~Wu, Z.~Zhang, G.~Ma, C.~Zhang, L.~Chen, A.~V. Balakin, A.~P. Shkurinov, Y.~Zhu, and S.~Zhuang, ``{Cascaded Amplification and Manipulation of Terahertz Emission by Flexible Spintronic Heterostructures},'' {\em Laser Photonics Rev.}, vol.~16, p.~2100688, Sept. 2022.

\bibitem{feng2021spintronic}
Z.~Feng, H.~Qiu, D.~Wang, C.~Zhang, S.~Sun, B.~Jin, and W.~Tan, ``Spintronic terahertz emitter,'' {\em Journal of Applied Physics}, vol.~129, no.~1, p.~010901, 2021.

\bibitem{papaioannou2020thz}
E.~T. Papaioannou and R.~Beigang, ``Thz spintronic emitters: a review on achievements and future challenges,'' {\em Nanophotonics}, vol.~10, no.~4, pp.~1243--1257, 2020.

\bibitem{godlike}
R.~Oldenhuis, ``{GODLIKE - A robust single-\& multi-objective optimizer}.'' GitHub, https://github.com/rodyo/FEX-GODLIKE/releases/tag/v1.5, 2023.

\bibitem{lagarias1998convergence}
J.~C. Lagarias, J.~A. Reeds, M.~H. Wright, and P.~E. Wright, ``{Convergence properties of the Nelder--Mead simplex method in low dimensions},'' {\em SIAM Journal on optimization}, vol.~9, no.~1, pp.~112--147, 1998.

\bibitem{rouzegar2023broadband}
R.~Rouzegar, A.~L. Chekhov, Y.~Behovits, B.~R. Serrano, M.~A. Syskaki, C.~H. Lambert, D.~Engel, U.~Martens, M.~M{\"u}nzenberg, M.~Wolf, {\em et~al.}, ``{Broadband Spintronic Terahertz Source with Peak Electric Fields Exceeding 1.5 MV/cm},'' {\em Physical Review Applied}, vol.~19, no.~3, p.~034018, 2023.

\bibitem{kroll2007metallic}
J.~Kr{\"o}ll, J.~Darmo, and K.~Unterrainer, ``Metallic wave-impedance matching layers for broadband terahertz optical systems,'' {\em Optics express}, vol.~15, no.~11, pp.~6552--6560, 2007.

\bibitem{ding2016ultrathin}
L.~Ding, X.~Wang, N.~S.~S. Ang, C.~Lu, V.~Suresh, S.-J. Chua, and J.~Teng, ``Ultrathin film broadband terahertz antireflection coating based on impedance matching method,'' {\em IEEE Journal of Selected Topics in Quantum Electronics}, vol.~23, no.~4, pp.~1--8, 2016.

\bibitem{carli1977reflectivity}
B.~Carli, ``Reflectivity of metallic films in the infrared,'' {\em JOSA}, vol.~67, no.~7, pp.~908--910, 1977.

\bibitem{Kampfrath_EOS_TF}
T.~Kampfrath, J.~Nötzold, and M.~Wolf, ``Sampling of broadband terahertz pulses with thick electro-optic crystals,'' {\em Applied Physics Letters}, vol.~90, 2007.

\bibitem{Battiato2010Jul}
M.~Battiato, K.~Carva, and P.~M. Oppeneer, ``{Superdiffusive Spin Transport as a Mechanism of Ultrafast Demagnetization},'' {\em Phys. Rev. Lett.}, vol.~105, p.~027203, July 2010.

\bibitem{Burghartz1994Sep}
{\relax St}.~Burghartz and B.~Schulz, ``{Thermophysical properties of sapphire, AlN and MgAl$_2$O$_4$ down to 70 K},'' {\em J. Nucl. Mater.}, vol.~212-215, pp.~1065--1068, Sept. 1994.

\bibitem{Shan2003Jun}
J.~Shan, F.~Wang, E.~Knoesel, M.~Bonn, and T.~F. Heinz, ``{Measurement of the Frequency-Dependent Conductivity in Sapphire},'' {\em Phys. Rev. Lett.}, vol.~90, p.~247401, June 2003.

\bibitem{Talghader2004Apr}
J.~J. Talghader, ``{Thermal and mechanical phenomena in micromechanical optics},'' {\em J. Phys. D: Appl. Phys.}, vol.~37, p.~R109, Apr. 2004.

\bibitem{Sinha1978Apr}
A.~K. Sinha, H.~J. Levinstein, and T.~E. Smith, ``{Thermal stresses and cracking resistance of dielectric films (SiN, Si$_3$N$_4$, and SiO$_2$) on Si substrates},'' {\em J. Appl. Phys.}, vol.~49, pp.~2423--2426, Apr. 1978.

\bibitem{Kumar2017Jun}
S.~Kumar, P.~S. Maji, and R.~Das, ``{Tamm-plasmon resonance based temperature sensor in a Ta$_2$O$_5$/SiO$_2$ based distributed Bragg reflector},'' {\em Sens. Actuators, A}, vol.~260, pp.~10--15, June 2017.

\bibitem{malitson1965interspecimen}
I.~H. Malitson, ``Interspecimen comparison of the refractive index of fused silica,'' {\em Josa}, vol.~55, no.~10, pp.~1205--1209, 1965.

\bibitem{luke2015broadband}
K.~Luke, Y.~Okawachi, M.~R. Lamont, A.~L. Gaeta, and M.~Lipson, ``{Broadband mid-infrared frequency comb generation in a Si$_3$N$_4$ microresonator},'' {\em Optics letters}, vol.~40, no.~21, pp.~4823--4826, 2015.

\bibitem{KaliteevskiPRB2007}
M.~Kaliteevski, I.~Iorsh, S.~Brand, R.~A. Abram, J.~M. Chamberlain, A.~V. Kavokin, and I.~A. Shelykh, ``{Tamm plasmon-polaritons: Possible electromagnetic states at the interface of a metal and a dielectric Bragg mirror},'' {\em Physical Review B}, vol.~76, p.~165415, 10 2007.

\bibitem{Jiang2023}
Y.~Jiang, Y.~Jiang, H.~Li, H.~Li, X.~Zhang, X.~Zhang, F.~Zhang, F.~Zhang, Y.~Xu, Y.~Xu, Y.~Xiao, F.~Liu, F.~Liu, A.~Wang, Q.~Zhan, W.~Zhao, and W.~Zhao, ``{Promoting spintronic terahertz radiation via Tamm plasmon coupling},'' {\em Photonics Res.}, vol.~11, pp.~1057--1066, June 2023.

\bibitem{Tao2018Jun}
X.~Tao, Q.~Liu, B.~Miao, R.~Yu, Z.~Feng, L.~Sun, B.~You, J.~Du, K.~Chen, S.~Zhang, L.~Zhang, Z.~Yuan, D.~Wu, and H.~Ding, ``{Self-consistent determination of spin Hall angle and spin diffusion length in Pt and Pd: The role of the interface spin loss},'' {\em Sci. Adv.}, vol.~4, June 2018.

\bibitem{Yu2018Jul}
R.~Yu, B.~F. Miao, L.~Sun, Q.~Liu, J.~Du, P.~Omelchenko, B.~Heinrich, M.~Wu, and H.~F. Ding, ``{Determination of spin Hall angle and spin diffusion length in $\ensuremath{\beta}$-phase-dominated tantalum},'' {\em Phys. Rev. Mater.}, vol.~2, p.~074406, July 2018.

\bibitem{Saleh1991Aug}
B.~E.~A. Saleh and M.~C. Teich, {\em {Fundamentals of Photonics}}.
\newblock John Wiley \& Sons, Inc, Aug. 1991.

\bibitem{Button1979}
K.~J. Button, {\em {Infrared and Millimeter Waves}}.
\newblock Walthm, MA, USA: Elsevier, 1979.

\bibitem{Marple1964Mar}
D.~T.~F. Marple, ``{Refractive Index of ZnSe, ZnTe, and CdTe},'' {\em J. Appl. Phys.}, vol.~35, pp.~539--542, Mar. 1964.

\bibitem{MGO}
``{Refractive index of MgO (Magnesium monoxide) - Stephens},'' Feb. 2024.
\newblock [Online; accessed 1. Feb. 2024].

\bibitem{Ghosh1999May}
G.~Ghosh, ``{Dispersion-equation coefficients for the refractive index and birefringence of calcite and quartz crystals},'' {\em Opt. Commun.}, vol.~163, pp.~95--102, May 1999.

\bibitem{Li1980Jan}
H.~H. Li, ``{Refractive index of alkaline earth halides and its wavelength and temperature derivatives},'' {\em J. Phys. Chem. Ref. Data}, vol.~9, pp.~161--290, Jan. 1980.

\bibitem{Phillip1964Nov}
H.~R. Phillip and E.~A. Taft, ``{Kramers-Kronig Analysis of Reflectance Data for Diamond},'' {\em Phys. Rev.}, vol.~136, pp.~A1445--A1448, Nov. 1964.

\bibitem{sapphire}
``{Refractive index of Al2O3 (Aluminium sesquioxide, Sapphire, Alumina) - Malitson-o},'' Feb. 2024.
\newblock [Online; accessed 1. Feb. 2024].

\bibitem{Tsuda2018Mar}
S.~Tsuda, S.~Yamaguchi, Y.~Kanamori, and H.~Yugami, ``{Spectral and angular shaping of infrared radiation in a polymer resonator with molecular vibrational modes},'' {\em Opt. Express}, vol.~26, pp.~6899--6915, Mar. 2018.

\bibitem{Rogalin2018Dec}
V.~E. Rogalin, I.~A. Kaplunov, and G.~I. Kropotov, ``{Optical Materials for the THz Range},'' {\em Opt. Spectrosc.}, vol.~125, pp.~1053--1064, Dec. 2018.

\bibitem{Sajadi2015Nov}
M.~Sajadi, M.~Wolf, and T.~Kampfrath, ``{Terahertz-field-induced optical birefringence in common window and substrate materials},'' {\em Opt. Express}, vol.~23, pp.~28985--28992, Nov. 2015.

\bibitem{Tcypkin2020Oct}
A.~N. Tcypkin, M.~Melnik, I.~O. Vorontsova, S.~A. Kozlov, and M.~O. Zhukova, ``{Estimations of Low-Inertia Cubic Nonlinearity Featured by Electro-Optical Crystals in the THz Range},'' {\em Photonics}, vol.~7, p.~98, Oct. 2020.

\bibitem{Grischkowsky1990Oct}
D.~Grischkowsky, S.~Keiding, M.~van Exter, and {\relax Ch}.~Fattinger, ``{Far-infrared time-domain spectroscopy with terahertz beams of dielectrics and semiconductors},'' {\em J. Opt. Soc. Am. B, JOSAB}, vol.~7, pp.~2006--2015, Oct. 1990.

\bibitem{Palik1997}
E.~D. Palik, {\em {Handbook of Optical Constants of Solids}}.
\newblock Elsevier, Academic Press, 1997.

\bibitem{Fan2015Oct}
F.~Fan, X.~Zhang, S.~Li, D.~Deng, N.~Wang, H.~Zhang, and S.~Chang, ``{Terahertz transmission and sensing properties of microstructured PMMA tube waveguide},'' {\em Opt. Express}, vol.~23, pp.~27204--27212, Oct. 2015.

\bibitem{Du2021Nov}
L.~Du, F.~Roeder, Y.~Li, M.~Shalaby, B.~Beleites, F.~Ronneberger, and A.~Gopal, ``{Organic crystal-based THz source for complex refractive index measurements of window materials using single-shot THz spectroscopy},'' {\em Appl. Phys. A}, vol.~127, pp.~1--11, Nov. 2021.

\bibitem{Islam2020May}
M.~S. Islam, C.~M.~B. Cordeiro, M.~J. Nine, J.~Sultana, A.~L.~S. Cruz, A.~Dinovitser, B.~W.-H. Ng, H.~Ebendorff-Heidepriem, D.~Losic, and D.~Abbott, ``{Experimental Study on Glass and Polymers: Determining the Optimal Material for Potential Use in Terahertz Technology},'' {\em IEEE Access}, vol.~8, pp.~97204--97214, May 2020.

\bibitem{nenno2019modification}
D.~M. Nenno, L.~Scheuer, D.~Sokoluk, S.~Keller, G.~Torosyan, A.~Brodyanski, J.~L{\"o}sch, M.~Battiato, M.~Rahm, R.~H. Binder, {\em et~al.}, ``Modification of spintronic terahertz emitter performance through defect engineering,'' {\em Scientific reports}, vol.~9, no.~1, pp.~1--16, 2019.

\end{thebibliography}

\end{document}